\documentclass[final,3p,times]{elsarticle}

\usepackage{amssymb}
\usepackage{graphicx}
\usepackage{epstopdf, epsfig}
\usepackage{placeins}
\usepackage{xcolor}
\usepackage{amsmath}
\usepackage[colorlinks,citecolor=red,urlcolor=blue,bookmarks=false]{hyperref}
\usepackage[capitalise]{cleveref}
\usepackage{flexisym}
\usepackage{booktabs} 
\usepackage{verbatim} 
\usepackage{algorithm}

\creflabelformat{equation}{#2#1#3}
\crefformat{appendix}{#2#1#3}

\newcommand{\gv}[1]{\ensuremath{\mbox{\boldmath$ #1 $}}} 
\newcommand{\bv}[1]{\ensuremath{\boldsymbol{#1}}} 
\newcommand{\norm}[1]{\left\lVert#1\right\rVert}
\newcommand{\imag}[1]{\mathrm{Im} \left[ #1 \right]}
\newcommand{\RN}[1]{\textup{\uppercase\expandafter{\romannumeral#1}}}
\newcommand{\order}[1]{\mathcal O \left( #1 \right)}
\newcommand{\vort}{\omega}
\newcommand{\pen}{_{\scriptscriptstyle \lambda}}
\newcommand{\Rey}{\ensuremath{Re}}
\newcommand{\Er}{\ensuremath{Er}}
\newcommand{\Ca}{\ensuremath{Cau}}
\newcommand{\LCFL}{\ensuremath{\textrm{LCFL}}}
\newcommand{\CFL}{\ensuremath{\textrm{CFL}}}
\newcommand{\refmap}{inverse map }
\newcommand{\Refmap}{Inverse map }
\newcommand{\dev}{\textprime}
\newcommand{\Ltwo}{\ensuremath{L_{2}} }
\newcommand{\Linf}{\ensuremath{L_{\infty}} }
\newcommand{\capsub}[1]{(\textit{#1})}
\newcommand{\todo}[1]{\textcolor{red}{TODO: #1}\PackageWarning{TODO:}{#1!}}

\journal{Elsevier}

\begin{document}

\begin{frontmatter}


\title{A remeshed vortex method for mixed rigid/soft body fluid–structure interaction}

\author[1]{Yashraj Bhosale \fnref{fn1}}
\author[1]{Tejaswin Parthasarathy\fnref{fn1}}
\author[1,2,3]{Mattia Gazzola \corref{cor1}} \ead{mgazzola@illinois.edu}
\cortext[cor1]{Corresponding author}
\fntext[fn1]{Equal contribution.}
\address[1]{Mechanical Sciences and Engineering, University of Illinois at
Urbana-Champaign, Urbana, IL 61801, USA}
\address[2]{National Center for Supercomputing Applications, University of Illinois at
Urbana-Champaign, Urbana, IL 61801, USA}
\address[3]{Carl R. Woese Institute for Genomic Biology, University of Illinois at
Urbana-Champaign, Urbana, IL 61801, USA}

\author{}

\address{}

\begin{abstract}
We outline a 2D algorithm for solving incompressible flow–structure interaction problems for
mixed rigid/soft body representations, within a consistent framework based on the remeshed
vortex method. We adopt the one-continuum formulation to represent both solid and fluid
phases on an Eulerian grid, separated by a diffuse interface. Rigid solids are treated
using Brinkman penalization while an inverse map technique is used to obtain elastic stresses
in the hyperelastic solid phase. We test our solver against a number of benchmark problems,
which demonstrate physical accuracy and first to second order convergence in space and time.
Benchmarks are complemented by additional investigations that illustrate the ability of
our numerical scheme to capture essential fluid–structure interaction phenomena across a
variety of scenarios involving internal muscular actuation, self propulsion, multi-body contact,
heat transfer and rectified viscous streaming effects. Through these illustrations, we
showcase the ability of our solver to robustly deal with different constitutive models
and boundary conditions, solve disparate multi-physics problems and achieve faster time-to-solutions
by sidestepping CFL time step restrictions.
\end{abstract}

\begin{graphicalabstract}
\begin{figure}[!ht]
\centering
\includegraphics[width=\textwidth]{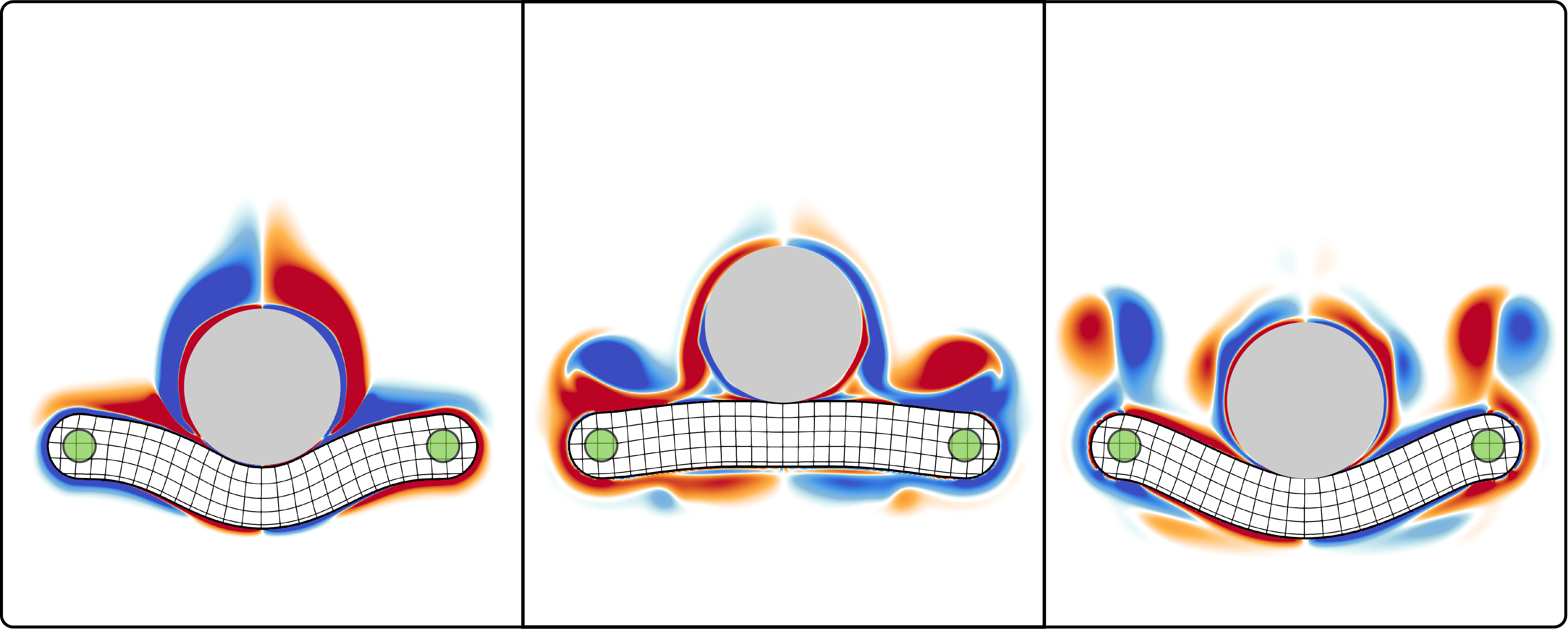}
\label{fig:gabs}
\end{figure}
\end{graphicalabstract}

\begin{highlights}
\item Unified formulation based on remeshed vortex method for the simulation of
	multiple, heterogeneous rigid/elastic body dynamics immersed in viscous fluids
\item Rigorous benchmarking and convergence analysis against a battery of
    theoretical/numerical tests
\item The proposed method demonstrates accuracy, robustness and versatility in dealing
	with a variety of boundary conditions, constitutive and actuation models,
	across multi-physics scenarios that include muscular actuation, self-propulsion,
	multi-body contact, heat transfer and rectified viscous streaming effects.
\item Sidestepping conventional CFL time-step restrictions for
    advection dominated problems
\item Stable simulations of purely elastic bodies without the need of internal
    damping/dissipation
\end{highlights}


\end{frontmatter}

\section{Introduction}\label{sec:intro}

This paper presents a remeshed vortex method based formulation that captures
essential two-way flow--structure interactions among multiple heterogeneous  soft and
rigid bodies immersed in an incompressible viscous fluid. We are motivated by the
relevance of these effects in engineering and biology across scales \cite{alben2002drag,
pozrikidis2003modeling, alben2004flexibility,tytell2016role,parthasarathy2017effect, goza2018global,
lagrone2019elastohydrodynamics, bhosale2020bending}, particularly in the context of soft
robotics \cite{park2016phototactic, zhang2019modeling, aydin2019neuromuscular} and
biolocomotion \cite{gazzola2014scaling, gazzola2015gait}, where there exists an
inextricable nexus between compliant mechanics, environmental interactions, control and
behaviour. Accurate and versatile solvers are then key to shed light on and dissect
underlying mechanisms and design solutions, with potential applications beyond the above
domains: from medicine, where compliant devices may be used to deliver drugs
\cite{ceylan2017mobile}, to inertial microfluidics, where streaming effects
\cite{lutz2005microscopic, marmottant2004bubble, liu2003hybridization}
can be used for particle manipulation, or flow control for drag reduction or heat
transfer enhancement. These problems are typically characterized by non-linear
interfacial-driven coupled dynamics across disparate solid and fluid spatio-temporal
scales, complex solid morphologies and boundary conditions, and large elastic deformations.
Because of these features, numerical studies in these settings have been
traditionally challenging, and discoveries 
have been predominantly (although not exclusively) led by experiments,
which are expensive and time consuming. Nonetheless, computational inroads have been made
over the years.
The resulting algorithms can be broadly classified into three major categories based on the
representation of the fluid and solid phases~\cite{jain2019conservative}. These entail
fully Lagrangian formulations, fully Eulerian formulations and mixed Lagrangian–Eulerian
formulations.

In fully Lagrangian formulations, both fluid and solid phases are represented as
particles advected by flow and solid velocity fields.
Also known as meshless methods, popular members of this class include the reproducing kernel
particle method (RKPM)
\cite{liu1995reproducing} and smoothed-particle hydrodynamics (SPH) \cite{gingold1977smoothed}.
These methods present a number of attractive features such as simplified
parallelization, good conservation properties and automatic local ($r$-) adaptivity~\cite{price2011smoothed}. At
the same time they are limited in their ability to treat boundary
conditions, are accompanied by higher algorithmic costs compared to traditional grid based methods
\cite{price2011smoothed,shadloo2016smoothed}, and may incur particle distortion effects
that can severely impair accuracy.

On the other side of the spectrum lies the fully Eulerian formulation in which both the phases are
represented on a fixed Eulerian grid or mesh, with the solid--fluid boundary usually tracked
using implicit techniques such as level set~\cite{sethian1999level} and volume of
fluid (VOF)~\cite{noh1976slic} method. This category has seen recent developments
through the use of reference map technique coupled with level sets
\cite{valkov2015eulerian}, and Cauchy--Green tensor advection coupled with VOF
\cite{sugiyama2011full}, for the treatment of elastic solids immersed in
viscous fluids. These approaches are typically implemented through finite differences
\cite{sugiyama2011full} or finite volumes \cite{jain2019conservative}.
These methods have been shown to successfully capture flow past elastic bodies,
self-propulsion, solid-solid contact physics, or haemodynamics among others
\cite{jain2019conservative, valkov2015eulerian, sugiyama2011full, sugiyama2010full, nagano2010full},
and offer a number of attractive features such as cost effectiveness due to the fixed mesh,
straightforward evaluation of operators and simplicity in parallelization. At the same time,
they are hampered by difficulties in resolving slender structures, treating far field boundary
conditions and face advection based, CFL time step restrictions.

Finally, the most commonly used, diverse and historically significant class is the mixed
Lagrangian--Eulerian formulation, where the solid phase is represented on a Lagrangian grid while
the fluid phase is usually represented on a fixed Eulerian mesh.  This class
can be divided into two major sub classes, namely
partitioned domain methods and monolithic domain methods. Partitioned domain methods are
characterized by separate meshes/solution spaces for the solid and fluid phases, and typically include
members such as arbitrary Lagrangian-Eulerian (ALE) approach \cite{hu2001direct} and
deforming-spatial-domain/stabilized-space-time approach (DST/SST) \cite{tezduyar1992new},
within the context of
finite element methods. These established methods, while mathematically involved, possess rigorous
convergence properties and have proven useful in a number of applications, from
(bio-)propulsion \cite{takizawa2012space} to cardiovascular
modeling~\cite{watanabe2004multiphysics} or aerodynamics~\cite{takizawa2011stabilized}. Yet,
their parallel implementation might be challenging given their partitioned approach,
while also requiring generation of a new grid at every time step to avoid computational element
distortion, both of which renders them computationally expensive for highly deforming elastic solids~\cite{hu1996direct}.
Monolithic domain methods
instead solve a single set of governing equations over the entire domain with the solid--fluid coupling
boundary conditions formulated as appropriate forcing terms. Well
known members include immersed methods~\cite{griffith2020immersed}
(immersed boundary~\cite{peskin2002immersed,
uhlmann2005immersed,taira2007immersed,wang2015strongly,goza2017strongly}, immersed finite
element \cite{zhang2004immersed,zhao2008fixed,tian2014fluid,griffith2017hybrid} and
immersed interface~\cite{li2001immersed} methods) and fictitious domain methods
\cite{glowinski2001fictitious,yu2005dlm,engels2015numerical}.
These methods are known for their versatility and have been widely used to study flow past
complex geometries~\cite{peskin2002immersed,uhlmann2005immersed}, bio-mimetic propulsion
\cite{wang2015strongly, zhao2008fixed, tian2014fluid, engels2015numerical,
eldredge2007numerical, eldredge2008dynamically}, hemodynamics
\cite{peskin2002immersed,griffith2017hybrid} and flow induced vibration
\cite{wang2015strongly, goza2017strongly, yu2005dlm, eldredge2008dynamically}. However,
these methods also face advection based CFL time step restrictions, as well as
difficulties in achieving higher-order convergence. We note that while the classification above
serves as a useful, high level guidance, methods might straddle
 across categories. For a more detailed classification, we refer the reader to the
 recent paper of~\citet{jain2019conservative}.

An alternative approach known as remeshed vortex method has developed considerably in the past
decade to mitigate advection time step restrictions, while offering high accuracy.
It represents the solid phase on a fixed Eulerian mesh,
while the fluid alternates between a Lagrangian and Eulerian description to solve for the
velocity--vorticity formulation of the momentum equation (as opposed to the velocity--pressure
formulation used in other methods)
\cite{beale1982vortex,leonard1985computing,raviart1985analysis,cottet2000vortex,winckelmans2004vortex}.
It carries over a number of attractive features from Lagrangian and Eulerian methods, including guaranteed convergence, stability, accuracy,
compact support of vorticity leading to automatic local ($r$-)adaptivity, natural treatment of
far field boundary conditions, ability to model complex solid morphologies, relaxed
advection stability conditions, and computational economy rivaling traditional grid based methods.
This makes it a versatile method to capture the presence of unsteady, complex,
deforming
bodies~\cite{gazzola2011simulations,gazzola2012c,van2013optimal,bernier2019simulations}
across scales~\cite{rossinelli2015mrag,Gazzola:2014,Gazzola:2012a},
to deal with contact physics~\cite{coquerelle2008vortex}, multiphase~\cite{lorieul2018development} and compressible flows~\cite{eldredge2002vortex,parmentier2018vortex}, in 2D as
well as in 3D~\cite{van2013optimal, winckelmans1993contributions,
ploumhans2000vortex}.
Yet, despite this versatility, little effort has been made to capitalize on these advantages
to solve the strongly coupled equations of motion between multiple heterogeneous soft and
rigid bodies and surrounding fluid.

In this work, we provide this crucial contribution.
Specifically, we combine inverse map technique and Brinkman penalization within a
consistent and seamless one-fluid formulation to account for full two-way coupling between
an incompressible viscous fluid and multiple, heterogeneous rigid and elastic bodies.
This approach combines the attractive features of vortex methods, with the ones of the inverse
map technique, namely, straightforward solid stress evaluation, stability and convenient
solid--fluid interface tracking based on the same machinery of the Brinkman
penalization. While previous attempts employed simplified 1D formulations
leveraging the slenderness of thin elastic structures \cite{goza2017strongly,
engels2013two}, our method solves for bulk elasticity and enables the simulation of
arbitrarily shaped 2D soft bodies. Through numerous benchmarks and illustrations, we then
demonstrate the accuracy, robustness and versatility of our solver across multiphysics
scenarios, boundary conditions, constitutive and actuation models. 

The work is organized as follows: governing equations and the various techniques used to
solve them are described in \cref{sec:gov} and \cref{sec:method}, respectively; the
proposed algorithm and the numerical discretization is detailed in \cref{sec:num};
rigorous benchmarking and convergence analysis is presented in \cref{sec:bmks};
versatility and robustness of the solver is illustrated through a variety of multifaceted
cases in \cref{sec:ills}; finally, concluding remarks are provided in \cref{sec:conc}.

\section{Governing equations}\label{sec:gov}

In this section, we present the complete set of governing equations and constitutive laws
that define the dynamics of multiple rigid/elastic bodies immersed in a viscous fluid.

\subsection{Governing equations for solids and fluids}\label{sec:cauchy_gov}

We consider a two-dimensional domain \(\Sigma\) physically occupied by a
viscous fluid and rigid and elastic bodies. We denote with \(\Omega_{e,i}\) \&
\(\partial\Omega_{e,i}, \; i=1,\dots,N_{e}\) and \(\Omega_{r,j}\) \&
\(\partial\Omega_{r,j}, \; j=1,\dots,N_{r}\) the support and boundaries of the elastic
and rigid solids, respectively. Denoting \(\overline{\Omega} = \overline{\Omega_{e,i}} \cup
  \overline{\Omega_{r,j}}\) to be the region occupied by solid material, the
fluid then occupies the region \(\Sigma - \overline{\Omega}\).

Linear and angular momentum balance of elastic solid and fluid domains
(for Eulerian differential volumes \( d\gv{x} \)), result in the Cauchy momentum equation

\begin{equation}
\label{eqn:cauchy1}
    \frac{\partial \gv{v}}{\partial t} + \bv{\nabla} \cdot
    \left( \gv{v} \bv{\otimes} \gv{v}\right) = -\frac{1}{\rho}\nabla p
    + \gv{b}
	+ \frac{1}{\rho} \bv{\nabla}
    \cdot \bv{\sigma}\dev ,~~~~\gv{x}\in\Sigma\setminus\Omega_{r,j}
\end{equation}
where \( t \in \mathbb{R}^+ \) represents time,
\(\gv{v} : \Sigma \times \mathbb{R}^+ \mapsto \mathbb{R}^2\) represents the velocity
field, \(\rho\) denotes material density, \(p : \Sigma \times \mathbb{R}^+
\mapsto \mathbb{R}\) represents the hydrostatic pressure field, \( \gv{b} : \Sigma
\times \mathbb{R}^+ \mapsto \mathbb{R}^2 \) represents a conservative
volumetric body force field and \(\bv{\sigma}\dev : \Sigma \times \mathbb{R}^+
\mapsto \mathbb{R}^2 \otimes \mathbb{R}^2\) is the deviatoric
Cauchy stress tensor field. As a convention, the prime symbol \(\dev\) on a tensor $\bv{A}$
denotes it is deviatoric, i.e.
\(\bv{A}\dev := \bv{A} - \dfrac{1}{2} {tr}(\bv{A}) \bv{I}\),
with \(\bv{I}\) representing the tensor identity and
\({tr}(\cdot)\) representing the trace operator. We assume that all
fields defined above are sufficiently smooth in time and space. Incompressibility of the
fluid and elastic domains is kinematically enforced through
\begin{equation}
\label{eqn:incomp}
    \nabla \cdot \gv{v} \equiv 0,~~~~\boldsymbol{x}\in\Sigma\setminus\Omega_{r,j}.
\end{equation}
The fluid and elastic solid phases interact exclusively via boundary conditions,
imposing continuity in velocities (no-slip) and traction forces at all fluid--elastic solid
interfaces

\begin{equation}
\label{eqn:elastic_bcs}
    \gv{v} = \gv{v}_f = \gv{v}_{e, i},~~~~
    \bv{\sigma}\dev_f \cdot \gv{n} = \bv{\sigma}\dev_{e, i} \cdot \gv{n} ,~~~~\gv{x}\in
    \partial\Omega_{e,i}
\end{equation}
where $\gv{n}$ denotes the unit outward normal vector at the interface \(\partial \Omega_{e,i}\).
Here $\gv{v}_f$ and $\gv{v}_{e, i}$ correspond to the interfacial velocities in the fluid and
\(i^{\textrm{\scriptsize th}}\) elastic body, respectively, while $\bv{\sigma}\dev_f$ and
$\bv{\sigma}\dev_{e, i}$ correspond to the interfacial Cauchy stress tensor in the fluid
and \(i^{\textrm{\scriptsize th}}\) elastic body, respectively.

In the region \(\Omega_{r,j}, \; j=1,\dots,N_{r}\) occupied by rigid solids, the velocities
are kinematically restricted
to rigid body modes of pure translation and rotation. Hence, all rigid bodies interact with
the fluid domain only via the no-slip boundary condition

\begin{equation}
\label{eqn:rigid_bcs}
\gv{v} = \gv{v}_f = \gv{v}_{r, j} = \underbrace{\gv{\dot{x}}_{\textrm{cmr}, j}}_{\textrm{translation}}
+ \underbrace{\dot \theta_{r, j} \times (\gv{x} - \gv{x}_{\textrm{cmr}, j})}_{\textrm{rotation}},
~~~~\gv{x}\in \partial\Omega_{r, j}
\end{equation}
where \( \gv{v}_{r, j} \) is the rigid velocity field, \( \gv{x}_{\textrm{cmr}, j} \)
is the center of mass (COM) position, and \( \theta_{r, j} \) is the angular orientation
about this COM of the \(j^{\textrm{\scriptsize th}}\) rigid body.

\subsection{Constitutive laws for fluid and elastic solids}\label{sec:closure}

To close the above set of equations (\cref{eqn:cauchy1,eqn:incomp,eqn:elastic_bcs,eqn:rigid_bcs}) and
determine the system dynamics, it is necessary to specify the form of internal material
stresses, i.e. their constitutive laws. Here, we
discuss specific modeling choices for the Cauchy stress tensor
\(\bv{\sigma}\dev\) of \cref{eqn:cauchy1}, across the different phases.

The fluid is assumed to be Newtonian, isotropic and incompressible with density \(\rho_f\), dynamic
viscosity \(\mu_f\) and kinematic viscosity \(\nu_f = {\mu_f} / {\rho_f}\).
As such, the Cauchy stress is comprised of the purely viscous term

\begin{equation}
\label{eqn:cons_fluid}
\bv{\sigma}_{f}{\dev} := 2 \mu_{f} {\bv{D}}\dev
\end{equation}
where \({\bv{D}}\dev\) is the strain rate tensor \(\frac{1}{2} \left ( \bv{\nabla}
   \gv{v} + \bv{\nabla} \gv{v}^T \right)\).

Next, we assume that the elastic solid is isotropic, incompressible, has constant density \(\rho_e\) and
exhibits both elastic and viscous (or \emph{visco-elastic}) behavior. Then the
deviatoric Cauchy stress can be modeled as

\begin{equation}
\label{eqn:cons_solid}
\bv{\sigma}_{e}{\dev} := 2 \mu_{e} {\bv{D}}\dev + {\bv{\sigma}}\dev_{he}
\end{equation}
where \(\mu_{e}\) represents the dynamic
viscosity of the solid material (indicative of internal damping effects) and
\({\bv{D}}\dev\) is the strain rate tensor. For convenience, we can also
define the kinematic viscosity of the solid \(\nu_e = {\mu_e} / {\rho_e}\).

The term \(\bv{\sigma}\dev_{he}\) is the
hyperelastic contribution to the solid stress tensor. We describe it here through the
generalized Mooney--Rivlin model \cite{sugiyama2011full, bower2009applied}, developed to
capture finite-strain elastomeric and biological tissue material responses. We then consider
an elastic solid in a convective coordinate system evolving with time \(t\). At
\(t = 0\), the solid is in its initial, stress-free configuration. A material point location
within the solid is denoted by \(\gv{X} \in \Omega^{0}_{e} \subset \mathbb{R}^2\). Due to external
or internal forces and couples, the solid displaces and distorts in
physical space \(\gv{x} \in \Omega_{e}(t) \subset \mathbb{R}^2\) for \(t > 0\).
Phenomenologically, the stress field \(\bv{\sigma}\dev_{he}\) is
a function of the displacement \( \gv{u} = \gv{x} - \gv{X} \)
(or equivalently strain) of a solid material point, and arises from the strain energy
density function \(W\) stored in the solid due to deformations. This is equivalent to
\(\bv{\sigma}\dev_{he}\) being only dependent on the deformation gradient
\(\bv{F} := {\partial \gv{x}} / {\partial \gv{X}} \) and not on \( \gv{x} \)
itself (intuitively, purely rigid body motions cause no stress). Galilean invariance
dictates that this dependence on \( \bv{F} \)
occurs only through the rotationally-invariant left \( \bv{B} := \bv{F}\bv{F}^{T}\) or
right \( \bv{C} := \bv{F}^{T}\bv{F}\) Cauchy--Green deformation tensors. Without loss of
generality, the strain energy density \(W\) can then be modeled as a function of $\bv{C}$
only

\begin{equation}
\label{eqn:strain_energy1}
W(\bv{C}) := c_{1}\left(\widetilde{\RN{1}}_{\bv{C}}-2\right) +
c_{2}\left(\widetilde{\RN{2}}_{\bv{C}}-2\right) +
c_{3}\left(\widetilde{\RN{1}}_{\bv{C}} - 2\right)^{2}
\end{equation}
where $c_1$, $c_2$ and $c_3$ are material constants, and $\widetilde{\RN{1}}_{\bv{C}}$
and $\widetilde{\RN{2}}_{\bv{C}}$ are the reduced invariants of $\bv{C}$ defined as

\begin{equation}
\label{eqn:strain_energy2}
    \widetilde{\RN{1}}_{\bv{C}} := \frac{\RN{1}_{\bv{C}}}{\RN{3}^{1/3}_{\bv{C}}} ~, ~~~
    \widetilde{\RN{2}}_{\bv{C}} := \frac{\RN{2}_{\bv{C}}}{\RN{3}^{2/3}_{\bv{C}}}
\end{equation}
through the matrix invariants

\begin{equation}
\label{eqn:strain_energy3}
\RN{1}_{\bv{C}} := {tr}(\bv{C}) ~, ~~~
\RN{2}_{\bv{C}} := \frac{1}{2} \left( \RN{1}^2_{\bv{C}} -
{tr}(\bv{C} \cdot \bv{C}) \right) ~, ~~~
\RN{3}_{\bv{C}} := {det}(\bv{C})
\end{equation}
with ${det}(\cdot)$ representing the determinant operator.
By combining \cref{eqn:strain_energy1,eqn:strain_energy2,eqn:strain_energy3} and recalling
that for incompressible hyperelastic materials

\begin{equation}
\label{eqn:solid_stress_pre}
\bv{\sigma}\dev_{he} = \left( 2 \bv{F} \frac{\partial
W(\bv{C})}{\partial \bv{C}} \bv{F}^{T} \right)^\dev
\end{equation}
the final expression for the Cauchy stress reduces to

\begin{equation}
\label{eqn:solid_stress}
    \bv{\sigma}\dev_{he} = \left( 2 \bv{F} \left[
c_1 \frac{\partial \widetilde{\RN{1}}_{\bv{C}}}{\partial \bv{C}} +
c_2 \frac{\partial \widetilde{\RN{2}}_{\bv{C}}}{\partial \bv{C}} +
c_3 (\widetilde{\RN{1}}_{\bv{C}} - 2) \frac{\partial \widetilde{\RN{1}}_{\bv{C}}}{\partial \bv{C}}
\right] \bv{F}^{T} \right)^\dev
= \left(2 c_{1} \bv{B}+2 c_{2}({tr}(\bv{B}) \bv{B}-\bv{B} \cdot
\bv{B})+4 c_{3}({tr}(\bv{B})-2) \bv{B}\right){\dev}.
\end{equation}

For small deformations, the coefficients \(2 \left(
c_1 + c_2 \right)\) represent \(G\), the shear modulus of the solid, and
\(c_3\) is loosely related to the bulk modulus (\(K\)) of the material.
Finally, if we set \( c_2  = c_3 = 0\) and \( 2c_{1} = G \) in \cref{eqn:solid_stress},
we recover the Cauchy stress corresponding to a neo-Hookean material

\begin{equation}
\label{eqn:neo_hookean_stress}
\begin{aligned}
\bv{\sigma}\dev_{he} &= G \bv{B}{\dev}.
\end{aligned}
\end{equation}

We note that the above linear relation between \( \bv{\sigma}\dev_{he} \) and \( \bv{B}{\dev} \) does not amount to a linear stress-strain response as in perfectly elastic materials, because
\( \bv{B} := \bv{F}\bv{F}^{T} \) contains strain non-linearities which account for Galilean invariance. Indeed, the neo-Hookean model has been developed to capture non-linear stress-strain behaviours,
but differently from the generalized Mooney--Rivlin model, it does so to a lesser degree
of accuracy and generality.
Nonetheless, due to its popularity and for comparison purposes we consider
here the neo-Hookean model as well.

\section{Methodology}\label{sec:method}

With the fundamental governing equations and boundary conditions established, we now
present the techniques used to solve these equations. Our approach builds upon the
method developed in \citet{gazzola2011simulations} for rigid
body flow--structure simulations, but crucially augments it to account for the full two-way
coupling between fluids, rigid and elastic bodies, in a seamless fashion. For this, we use
the inverse map technique to track
solid deformations, couple it with a hyperelastic constitutive model and adopt the one
continuum formulation to solve the coupling problem in a unified remeshed vortex methods framework.

\subsection{Remeshed vortex method}\label{sec:rvm}

We consider the velocity--vorticity formulation of the 2D Cauchy momentum equation~\cref{eqn:cauchy1}
\begin{equation}
\label{eqn:simplifiedNS}
\frac{\partial \omega}{\partial t} + \nabla \cdot \left( \gv{v} \omega \right)
    =
    \underbrace{
    -\frac{{\nabla} \rho}{\rho^2} \times {\nabla} p
    + \frac{1}{\rho} {\nabla} \times \bv{\nabla} \cdot \bv{\sigma}\dev
    + {\nabla} \times \gv{b}}_{\textrm{RHS}}
\end{equation}
where $\omega : \Sigma \times \mathbb{R}^{+} \mapsto \mathbb{R} := {\nabla} \times \gv{v}$
represents the vorticity field.
Vortex methods discretize $\vort$ by means of particles,
characterized by their position $\gv{x}_p$, volume $V_p$ and strength corresponding to the
vorticity integral
$\Gamma_p = \int_{V_p} \vort d\mathbf{x}$. The advection of particles and quantities they
represent is performed in a
Lagrangian fashion where they move according to the velocity field $\gv{v}$
with strengths \( \Gamma_p\) evolving in accordance with RHS of \cref{eqn:simplifiedNS}.
\begin{equation}
\label{eqn:particles}
\frac{d\gv{x}_p}{dt} = \gv{v}(\gv{x}_p, t); ~~~~
\frac{d\Gamma_p}{dt} = \left[ \textrm{RHS} \right]_{V_p}
\end{equation}
In order to avoid Lagrangian distortion \cite{koumoutsakos2005multiscale}, a remeshing
approach is used. Particle strengths and locations are interpolated onto an underlying
regular grid at the end of each step using a high order, moment preserving interpolation
scheme \cite{gazzola2011simulations}.
This approach enables a number of favorable features: use of fast differential operators
to evaluate RHS terms, use of efficient Fourier transforms for solving Poisson equations,
numerical accuracy, relaxed stability condition for advection, compact vorticity support
and software scalability \cite{gazzola2011simulations,
rossinelli2015mrag, lorieul2018development,winckelmans1993contributions,
ploumhans2000vortex,rossinelli2010gpu}.

\subsection{Eulerian representation of interfaces using level sets}\label{sec:level_set}

All fluid--solid interfaces in our algorithm \(\partial\Omega_{i}\) are captured using
separate level set \cite{sethian1999level} functions
\( \phi_i : \Sigma \times \mathbb{R}^{+} \mapsto \mathbb{R} \) such that
\[ \partial\Omega_i = \{ \gv{x} \in \Sigma \mid \phi_i\left( \gv{x} , t\right) = 0 \}. \]
These interfaces are then advected by the velocity field \( \gv{v}(\gv{x}, t) \)
\begin{equation}
\frac{d\phi_i}{dt} + \gv{v} \cdot \nabla \phi_i = 0
\label{eq:phi_evolution}
\end{equation}
starting from their initial location
\( \phi_i\left( \gv{x} , 0 \right) = \phi^0_{i}(\gv{x}) \), with \( \phi^0_{i} \)
being a signed-distance function at time $t = 0$. The outward normal at the interface
is computed \cite{sethian1999level} as \( \gv{n}_i = \nabla \phi_i / \norm{\nabla \phi_i} \).

\subsection{Brinkman penalization}\label{sec:brinkman}

In order to account for the presence of rigid bodies, we employ the Brinkman penalization
technique \cite{carbou2003boundary, angot1999penalization}. In the
penalization technique, the flow velocity field is extended inside the rigid bodies,
and the Cauchy momentum equation (\cref{eqn:cauchy1}) is equipped with an additional forcing
term, to approximate the no-slip boundary conditions of \cref{eqn:rigid_bcs}
(see~\cite{bost2010convergence} for detailed proofs).

\begin{equation}
\label{eqn:cauchy2}
    \frac{\partial \gv{v}\pen}{\partial t} + \bv{\nabla} \cdot
    \left( \gv{v}\pen \bv{\otimes} \gv{v}\pen \right) =
    -\frac{1}{\rho}\nabla p\pen + \frac{1}{\rho} \bv{\nabla}
    \cdot \bv{\sigma}\dev\pen + \gv{b} + \lambda \sum_{i} H(\phi_{r, j}) (\gv{v}_{r, j} - \gv{v}),
    ~~~~ \nabla \cdot \gv{v}_{\lambda} = 0,~~~~\gv{x}\in\Sigma
\end{equation}
where $\lambda \gg 1$ is the penalization factor, $H(\cdot)$ denotes the Heaviside
function, $\phi_{r, j}$ corresponds to the level set which captures the interface of the
\(j^{\textrm{\scriptsize th}}\) rigid body and a subscript \( \lambda \) denotes the
 penalized fields satisfying the Brinkman--Cauchy~\cref{eqn:cauchy2}. This penalization factor $\lambda$ can be
chosen arbitrarily and directly controls the error in the penalized solution, bounded by
$\norm{\mathbf{v}-\mathbf{v}_{\lambda}} \le C\lambda^{-1/2}\norm{\mathbf{v}}$ \cite{carbou2003boundary}.
For a detailed discussion, the reader is referred to \cite{gazzola2011simulations}.

\subsection{Projection approach}\label{sec:projection}

While the no-slip condition is enforced via penalization, the feedback from the fluid
to the rigid bodies is captured using a projection approach and Newton’s equations
of motion

\begin{equation}
    \label{eqn:newton}
    m_{r, j} \ddot{\gv{x}}_{cmr, j} = \gv{F}^H_{r, j}~; ~~~ J_{r, j} \ddot{\theta}_{r, j} = M^H_{r, j}
\end{equation}
where  $m_{r, j}$, $J_{r, j}$, $\gv{F}^H_{r, j}$ and $M^H_{r, j}$ are,
respectively,  mass and moment of inertia of the \(j^{\textrm{\scriptsize th}}\) rigid
body, and hydrodynamic force and moment acting on it. At the start
of each time step, the flow is let to evolve freely over the entire domain as if the rigid
bodies were not there (i.e. the velocity field is evolved inside the bodies themselves).
The resulting new velocity field violates the rigid motion of the body, as well as its
no-slip condition. To recover correct motion and physical consistency, we project the evolved velocity
onto a subspace comprising only of rigid (translational and rotational) modes. Such
a projection is possible because the extra momentum flux that the body obtains from the
freely evolved flow correctly captures the feedback from the fluid onto the
body over the time step. After the rigid components of motion are recovered, they are used
to penalise the velocity field, thus regaining physical consistency, and to advect the
level sets. Therefore the interplay between projection and penalization allows to achieve
flow--structure coupling without the explicit use of forces and torques.
A detailed proof can be found in \cite{Patankar:2005}. We
conclude this section by noting that, in this case, the level set
advection equation \cref{eq:phi_evolution} can be semi-analytically solved, so as to
directly impose
\[ \phi_{r, j}(\gv{x}, t) = \phi^{0}_i(\gv{x}; \gv{x}_{cmr, j}(t), \theta_{r, j}(t)).\]

\subsection{Inverse map technique}\label{sec:invmap}

To capture the elastic solid phase dynamics, we need to compute the
deformation gradient tensor $\bv{F}$ in time. For this, two approaches may be used:
advect the Lagrangian tensor \(\bv{F}\) directly on a fixed grid or
\emph{remember} the material point of origin \(\gv{X}\) for all points
in the current solid phase \( \gv{x} \) and then compute the deformation gradient per
\(\bv{F} := {\partial \gv{x}} / {\partial \gv{X}}\). We choose the second
approach, and adopt the \refmap technique described below to
compute \(\bv{F}\) in a purely Eulerian fashion---for a detailed comparison
between these approaches the reader is referred to \cite{kamrin2009eulerian}.
This methodology has been (re)discovered many times across different communities
\cite{kamrin2009eulerian, belytschko2013nonlinear,pons2006maintaining,koopman2008numerical,
cottet2008eulerian, levin2011eulerian} and is
known by several names (inverse map \cite{belytschko2013nonlinear}, initial-point set
\cite{dunne2006eulerian}, LSPC \cite{pons2006maintaining}, original-coordinates
\cite{koopman2008numerical}, backward-characteristics \cite{milcent2016eulerian},
reference-map \cite{kamrin2009eulerian}, reference-coordinates \cite{levin2011eulerian}).
In the context of flow--structure interaction, it has found use in simulating elastic membranes
submerged in incompressible flow
\cite{cottet2008eulerian,milcent2016eulerian}, and recently it has
been extended to incompressible two-dimensional solids,
using the p--\gv{v} formulation of the Navier--Stokes equation and finite volumes and
differences \cite{jain2019conservative, rycroft2018reference}.

\begin{figure}[!ht]
\centering
\includegraphics[width=0.65\textwidth]{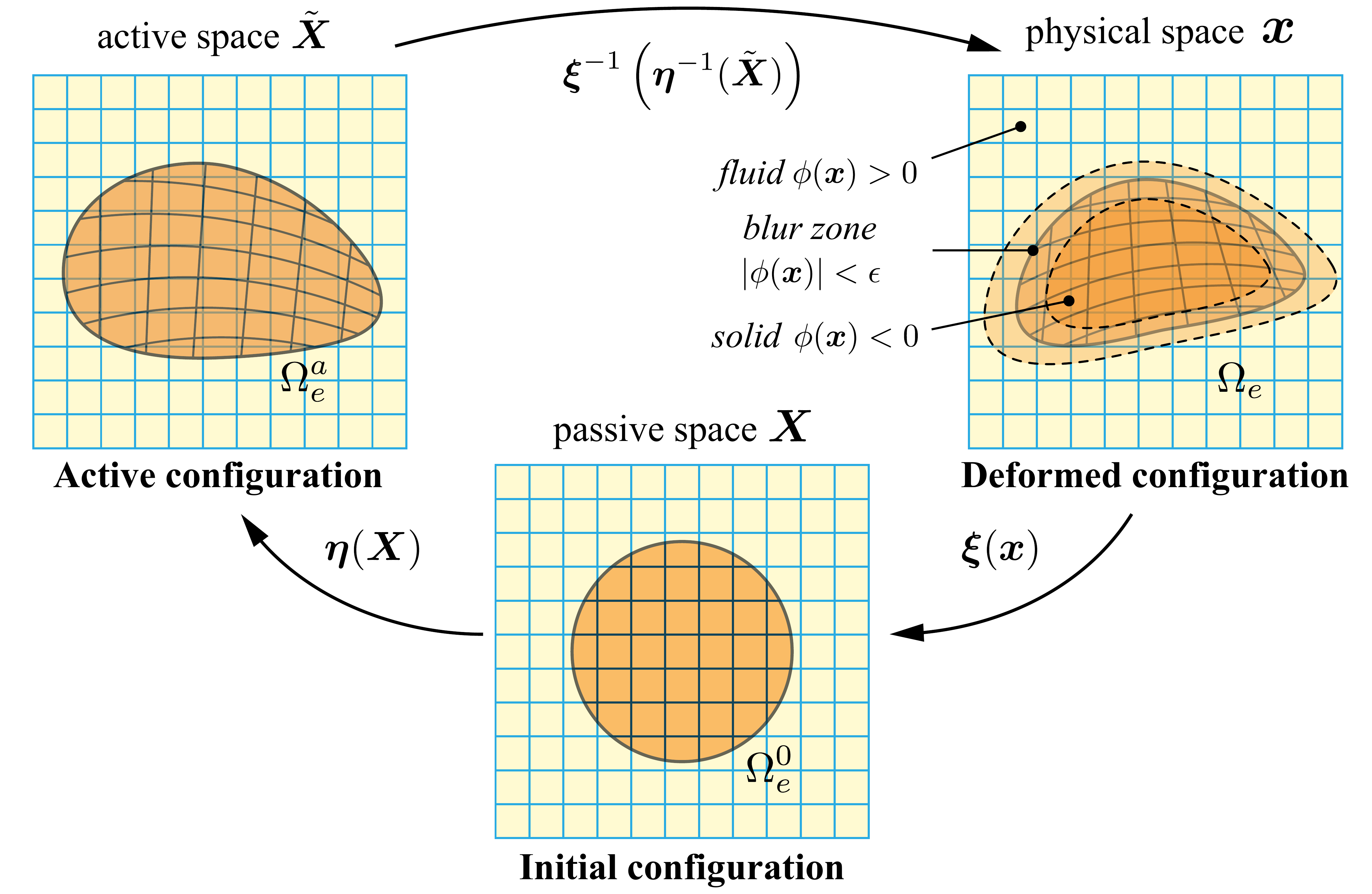}
\caption{
\label{fig:invmap}
Schematic of a deforming elastic solid, showing the initial (\(\Omega^{0}_{e}\),
represented by the passive space \(\gv{X}\)), deformed (\(\Omega_{e}\), represented
by the physical space \(\gv{x}\)) and active (\(\Omega^{a}_{e}\), represented by the
active space \(\tilde{\gv{X}}\))
configurations and their mappings \(\gv{\xi}\) and \(\gv{\eta }\). These
bijective maps allows us to transform between the three spaces (and hence their
respective configurations). Yellow squares (with blue borders) indicate the
background (discrete) Eulerian grid occupied by the fluid phase. Orange (with
black lines) indicate the Lagrangian grid of the solid, which we project onto
the background Eulerian grid. Additionally, upon discretization, the zero level
set contour ( \(\phi \left(\gv{x} \right) = 0\) ) of the solid is used to
distinguish the fluid \(\phi \left(\gv{x} \right) > 0\) and solid \(\phi
\left(\gv{x} \right) < 0\) phases, with mixed solid--fluid behavior in the
blur-zone \(\lvert \phi \left(\gv{x} \right) \rvert < \epsilon\).
}
\end{figure}

To illustrate the inverse map technique, we first consider an elastic solid in a convective
coordinate system (\cref{fig:invmap}), evolving with time \(t\). We start at
\(t = 0\) with the solid in its initial configuration, and denote a material point
within the solid by \(\gv{X} \in \Omega^{0}_{e} \subset \mathbb{R}^2\). Due to external
forces, the solid displaces and distorts occupying the
physical space \(\gv{x} \in \Omega_{e}(t) \subset \mathbb{R}^2\) at $t > 0$. Because of material
conservation, each point in \(\Omega\)\textsubscript{e} must have originated
from a univocal point in \(\Omega\)\textsuperscript{0}\textsubscript{e}, i.e. there must
exist a mapping \(\gv{\xi} : \Omega_{e} \times \mathbb{R}^{+} \mapsto \Omega^{0}_{e}\) such that
\( \gv{\xi}(\gv{x}, t):= \gv{X}  \)
with \(\gv{\xi}\) being sufficiently smooth (at least \(\mathcal{C}^{1}\) continuous),
and bijective. This diffeomorphic mapping is referred to as the inverse map. Physically, it denotes the
origin of the material point occupying Eulerian position \(\gv{x}\) at time \(t\).
From the definition above, \(\gv{\xi}\) is invariant for a material point (its origin is
always the same), implying that the material derivative of $\gv{\xi}$ is identically zero.
For an incompressible medium, this yields
\begin{equation}
\frac{\partial \gv{\xi}}{\partial t} + \gv{v} \cdot \bv{\nabla} \gv{\xi} = \gv{0} \; , \quad
\gv{\xi} (\gv{x},0) = \gv{x} = \gv{X}
\label{eq:inv_map_evolution}
\end{equation}
Therefore, the origin of a material point can be remembered as a field variable
governed by a pure advection evolution law.

The inverse map enables the computation of solid stresses in a straightforward
manner via \(\bv{F}\) (\cref{sec:closure}). Since \(\gv{\xi}(\gv{x}, t):= \gv{X}\), then
\({\partial \gv{X}} / {\partial \gv{x}} = \bv{\nabla}\gv{\xi}\) and hence
\(\bv{F} := {\partial \gv{x}} / {\partial \gv{X}} = \left( \bv{\nabla}\gv{\xi}
\right)^{-1}\), where the gradient \(\bv{\nabla}\) is a purely
Eulerian operator in physical space. Here the existence of \(\left( \bv{\nabla}\gv{\xi}
\right)^{-1}\) assumes bijectivity of \(\gv{\xi} \; \forall \; t > 0\), i.e. the
inverse map does not \emph{fold} over itself. Since the fluid zone is characterized by
high shear rates \( \norm{\gv{v}} \), \( \gv{\xi} \) may fold over.
To reduce this risk we only define \( \gv{\xi} \) inside the solid phase which has
characteristic low \( \norm{\gv{v}} \) values for any physical choice of \( W \).
In all our numerical simulations, we found that this choice \emph{prevented}
\( \gv{\xi} \) to fold and preserved its bijectivity.

An elastic solid material may undergo plastic effects or may be activated internally
(mimicking the effect of muscles \cite{nardinocchi2007active,Fan:2014}). In this case,
one can define an additional \emph{active} configuration
\(\Omega^{a}_{e}(t) \subset \mathbb{R}^2\) (\cref{fig:invmap}) that the solid tries to
approach to minimize its internal strain energy. We then
define this active configuration and introduce an additional diffeomorphic mapping
\(\gv{\eta} : \Omega^{0}_{e} \times \mathbb{R}^{+} \mapsto \Omega^{a}_{e}\) such
that \( \gv{\eta}\left( \gv{X}, t\right) := \tilde{\gv{X}} \)
where \(\tilde{\gv{X}}\) indicates a material point \(\in \Omega^{a}_{e}\).
Here \(\gv{\eta}\) can be directly specified (in the case of muscular activation) or
evolved separately under its own  specifics (such as in elasto-plasticity).
By composition of diffeomorphisms, we can obtain another diffeomorphic mapping relating
active and physical space \(\tilde{\gv{X}} =
\gv{\eta}\left( \gv{\xi}\left( \gv{x}, t\right), t \right) \Longleftrightarrow
\gv{x} = \gv{\xi}^{-1}\left( \gv{\eta}^{-1} \left( \tilde{\gv{X}}, t\right), t
\right)\). Then the total deformation gradient \(\bv{F}\) is \({\partial \gv{x}} / {\partial \tilde{\gv{X}}} =
{\partial \gv{x}} / {\partial {\gv{X}}} \cdot {\partial {\gv{X}}} / {\partial \tilde{\gv{X}}} =
\left(\bv{\nabla}\gv{\xi}\right)^{-1} \cdot
\left(\bv{\nabla}\gv{\eta}\right)^{-1}  = \left(\bv{\nabla}\gv{\eta} \cdot
\bv{\nabla}\gv{\xi}\right)^{-1}\) is fed into the constitutive
model (\cref{eqn:solid_stress}). This representation leads to a neatly compartmentalised
machinery, in which a variety of effects can be nested. We will demonstrate its use in
\cref{sec:jellyfish}, to simulate self-propelled, active and soft swimmers.

We equip each elastic body \( i \) with its own \( \gv{\xi}_i \) field.
This field can then be used to detect the interface \( \partial \Omega_{e, i} \), by
simply substituting \cref{eq:inv_map_evolution} in \cref{eq:phi_evolution} to obtain
\[ \phi_{e, i}(\gv{x}, t) = \phi^{0}_i(\gv{\xi}_i(\gv{x}, t)),\]
which is beneficial as we now do not need to evolve \( \phi \) in time, preserving
consistent interface positions between \( \gv{\xi}_i \) and \( \phi_i \) at all times.
As a final remark, we note that for all incompressible elastic materials
\(\det(\bv{F}) \equiv \det(\bv{\nabla}\gv{\xi}) \equiv \det(\bv{\nabla}\gv{\eta}) \equiv
1\). In our case this is identically satisfied as a byproduct of the
velocity field incompressibility (\cref{eqn:incomp}, see \citet{jain2019conservative} for a proof).

\subsection{Solid--fluid representation}\label{sec:rep}

With well defined governing equations, boundary conditions, constitutive laws
and interface characterization, we now proceed to describe the solid--fluid
representation used in our algorithm. To solve the
coupling problem, we adopt a conservative mixture model based on the one-fluid formulation
used in two-phase flows, also known as the \emph{one-continuum} formulation \cite{kataoka1986local}. In this
formulation, both solids and fluid share the same solution space and a monolithic velocity
field (see \cref{fig:invmap}). In the elastic solid regions, the Cauchy stress is computed using the solid
constitutive law (\cref{eqn:cons_solid}), while in the fluid zone the stress is computed
using the fluid constitutive law (\cref{eqn:cons_fluid}). Then, a Heaviside function
is used to smoothly blend the stresses and compute the monolithic Cauchy stress

\begin{equation}
    \label{eqn:blend_stress}
        \bv{\sigma}\dev = \sum_i H(\phi_{e, i}) ~ \bv{\sigma}\dev_{e, i} + \left( 1 - \sum_i
        H(\phi_{e, i}) \right) \bv{\sigma}\dev_f
\end{equation}
%
%
where $\sigma\dev_{e, i}$ and $\phi_{e, i}$ are, respectively, the solid stress tensor and
level set (defining the geometry) of the \(i^{\textrm{\scriptsize th}}\)
elastic body. Similarly, one can define a monolithic density field

\begin{equation}
    \label{eqn:blend_rho}
        \rho = \sum_i H(\phi_{e, i}) ~ \rho_{e, i} + \sum_j H(\phi_{r, j}) ~ \rho_{r, j} + \left( 1 - \sum_i
        H(\phi_{e, i}) - \sum_j H(\phi_{r, j}) \right) \rho_f
\end{equation}
where $\rho_{e, i}$ and $\rho_{r, j}$ represent the
density of the  \(i^{\textrm{\scriptsize th}}\) elastic body and the \(j^{\textrm{\scriptsize
th}}\) rigid body, respectively. Finally we note that the above formulation implicitly
satisfies the boundary conditions at the interface (\cref{eqn:elastic_bcs}), and allows for
the convenient use of common operators on the same solution space, across all the phases.

\subsection{Body and contact forces}\label{sec:coll}

Effects of external bulk forcing such as gravity can be directly captured through an
additional body force term $\gv{b} = \gv{g}$, where $\gv{g}$ is the
acceleration due to gravity. Additionally, in certain situations, bodies might
approach each other closely. In such cases we add to the Cauchy momentum equation
an extra contact forcing term that pushes these objects apart, to prevent their
interpenetration. Accordingly, we adopt the level set based contact
forcing model described in \citet{valkov2015eulerian}

\begin{equation}
    \label{eqn:coll}
    \gv{b}_{\textrm{coll}, i, j} =
\left\{ \begin{array}{lr}
\displaystyle k_{\textrm{coll}} \delta \left( \frac{\phi_i -
    \phi_j}{2} \right) \gv{n}_{i, j} &\mbox{$\phi_i < 0 ~ \textrm{or} ~ \phi_j < 0$}\\
\displaystyle 0 &\mbox{otherwise}
\end{array} \right.
\end{equation}
Here $k_{\textrm{coll}}$ is a constant while $\phi_i$ and $\phi_j$ correspond to the level
sets capturing the interface of the two bodies. The symbol $\delta(\cdot)$ stands for the Dirac Delta
function while $\gv{n}_{i, j}$ is a unit vector normal to the level sets of $\phi_i -
\phi_j$ and pointing away from the midplane level set contour, where $\phi_i = \phi_j$.

\subsection{Form of the Cauchy momentum equation to be numerically implemented}\label{sec:finalCauchy}

Here, we finally present the form of the Cauchy momentum equation that is ultimately
discretized and numerically implemented. Taking the curl of
\cref{eqn:cauchy2}, we obtain the vorticity formulation

\begin{equation}
\label{eqn:finalNS1}
\frac{\partial \omega}{\partial t} + \left( \gv{v} \cdot {\nabla} \right) \omega
    = \left( \omega \cdot \gv{\nabla} \right) \gv{v}
    -\frac{{\nabla} \rho}{\rho^2} \times \nabla p +
    + \frac{1}{\rho} {\nabla} \times \bv{\nabla} \cdot \bv{\sigma}\dev + {\nabla}
    \times \gv{b} + {\nabla} \times \lambda \sum_{j} H(\phi_{r, j}) (\gv{v}_{r, j} -
    \gv{v}).
\end{equation}
We then expand the baroclinic term as a function of the velocity and, considering the
fact that the stretching term $\left( \omega \cdot \gv{\nabla} \right) \gv{v}$
vanishes in two dimensions, we rewrite this equation as

\begin{equation}
\label{eqn:finalNS2}
    \frac{\partial \omega}{\partial t} + \left( \gv{v} \cdot \nabla \right) \omega =
    -\frac{{\nabla} \rho}{\rho} \times \left(\frac{\partial \gv{v}}{\partial t} +
    \bv{\nabla} \cdot \left( \gv{v} \otimes \gv{v} \right) - \gv{b} \right)
    + \frac{1}{\rho} \nabla \times \bv{\nabla} \cdot \bv{\sigma}\dev + \nabla
    \times \gv{b} + \nabla \times \lambda \sum_{j} H(\phi_{r, j}) (\gv{v}_{r, j} -
    \gv{v}).
\end{equation}
With pressure $p$ eliminated from the governing equations, an incompressible velocity
field is then directly recovered from the vorticity by solving a Poisson equation using
appropriate boundary conditions on \( \Sigma \)

\begin{equation}
\label{eqn:poisson}
    \nabla^2 \psi = - \omega; ~~~~
    \gv{v} = \nabla \times \psi
\end{equation}
where $\psi : \Sigma \times \mathbb{R}^+ \mapsto \mathbb{R}$ corresponds to the
streamfunction. In the next section, we proceed
to describe the numerical discretization of the elements
described above, with a detailed step by step explanation of the algorithm.

\section{Numerical discretization and algorithm}\label{sec:num}

We proceed to spatially discretize the system of equations (\cref{eqn:finalNS2,eqn:poisson}) by adopting a
Cartesian grid of uniform spacing $h$ which forms our computational domain \(\Sigma_{h}\).
All fields defined earlier are replaced by their discrete counterparts, now defined
on this discrete domain \(\Sigma_{h}\).
The temporal discretization is achieved via a Godunov split of~\cref{eqn:finalNS2}, which
leads to the algorithmic steps detailed in~\cref{alg}. This splitting enables us to
evaluate each step independently, providing the flexibility to conveniently mix explicit
and implicit time integration (\cref{v_pen} and
\cref{update_stress,update_baro,update_body,vort_advect,vort_reset,cm_update,ang_update}), at the
penalty of reducing convergence in time between first and second order (\cref{sec:bmks},
\cite{gazzola2011simulations}).
In the following, we describe one full time step of the proposed algorithm, from $t^n$ to
$t^{n + 1}$, assuming that all necessary quantities are known up to $t^n$.

\subsection{Poisson solve and velocity recovery}

We solve the Poisson ~\cref{poisson_solve} on the grid for periodic and unbounded boundary conditions
 \(\psi\) using a Fourier-series based \( \order{n \log(n)} \)
solver. This allows us to exploit the diagonality of the Poisson operator in the case of
periodic boundaries \cite{hockneycomputer} to achieve spectral accuracy.
For unbounded conditions, we use the zero padding technique of
\citet{hockneycomputer}, while for mixed periodic--unbounded boundaries we use the
approach of \citet{chatelain2010fourier}. Once \( \psi \) is obtained on the
grid, we recover the velocity per~\cref{vel_from_psi}, through the discrete second order
centered finite difference curl operator.

\begin{algorithm}
\caption{General algorithm}
\label{alg}
\begin{align}
    \textrm{Poisson solve} ~~~ &\nabla^2 \psi^n = - \omega^n
    \label{poisson_solve} \\
    \textrm{Velocity recovery} ~~~ &\gv{v}^n = \gv{\nabla} \times \psi^n + \gv{V}_{\infty}^n
    \label{vel_from_psi} \\
    \textrm{Rigid body level set recovery} ~~~ &\phi_{r, j}^n = \phi_{r, j}(\gv{x}_{\textrm{cmr}, j}^n, \theta_{r, j}^n, t^n)
    \label{chi_fix} \\
    \textrm{Translational projection} ~~~ &\gv{\dot{x}}_{\textrm{cmr}, j}^n = \dfrac{1}{M_{r, j}} \displaystyle \int_{\Sigma} \rho_{r, j}
    H_{\varepsilon}(\phi^n_{r, j}) \gv{v}^n
    d\gv{x}
    \label{lin_project} \\
    \textrm{Rotational projection} ~~~ &\gv{\dot \theta}_{r, j}^n = \dfrac{1}{J_{r, j}} \displaystyle \int_{\Sigma} \rho_{r,
    j}
    H_{\varepsilon}(\phi^n_{r, j}) \left[(\gv{x} - \gv{x}_{\textrm{cmr}, j}^n)
    \times \gv{v}^n\right]
    d\gv{x}
    \label{ang_project} \\
    \textrm{Rigid body velocity recovery} ~~~ &\gv{v}_{r, j}^n = \gv{\dot{x}}_{\textrm{cmr}, j}^n + \gv{\dot \theta}_{r, j}^n \times (\gv{x} -
    \gv{x}_{\textrm{cmr}, j}^n)
    \label{fin_project} \\
    \textrm{Velocity penalization} ~~~ &\gv{v}\pen^n = \frac{\gv{v}^n + \lambda \Delta t^n \sum_{j}
    H_{\varepsilon}(\phi_{r, j}) \gv{v}_{r, j}^n}
    {1 + \lambda \Delta t^n \sum_{j} H_{\varepsilon}(\phi_{r, j})}
    \label{v_pen} \\
    \textrm{Vorticity penalization} ~~~ &\omega\pen^n = \omega^n + \gv{\nabla} \times (\gv{v}\pen^n - \gv{v}^n)
    \label{vort_pen} \\
    \textrm{Inverse map advection} ~~~ &\dfrac{\partial \gv{\xi}^n}{\partial t}+\gv{v}_{\lambda}^n \cdot \nabla \gv{\xi}^n = 0
    \label{xi_advect} \\
    \textrm{Inverse map based level set recovery} ~~~ &\phi_{e, i}^n = \phi_{e, i}^0 (\gv{\xi}^n)
    \label{pin_phi} \\
    \textrm{Level set reinitialization} ~~~ & \lvert \gv{\nabla} \phi_{e, i}^n \rvert = 1
    \label{reinit_phi} \\
    \textrm{Inverse map extrapolation} ~~~ &\gv{\xi}_{\Sigma, ~ \textrm{extrap}}^n \leftarrow \gv{\xi}_{\Omega_e}^n
    \label{extrap_eta} \\
    \textrm{Monolithic stress computation} ~~~ &{\bv{\sigma}\dev}^{~n} = \bv{\sigma}\dev(\bv{\sigma}\dev_f^{~n}, \bv{\sigma}\dev_e^{~n})
    \label{blend_stress} \\
    \textrm{Monolithic density computation} ~~~ &\rho^n = \rho(\phi_{e, i}^n, ~ \rho_{e, i}, ~ \phi_{r, j}^n, ~ \rho_{r, j}, ~ \rho_f)
    \label{blend_rho} \\
    \textrm{Stress based vorticity update} ~~~ &\frac{\partial \omega\pen^n}{\partial t} = \frac{1}{\rho^n} \left(\gv{\nabla}
    \times \gv{\nabla} \cdot \bv{\sigma}\dev^{,n} \right)
    \label{update_stress} \\
    \textrm{Baroclinic term based vorticity update} ~~~ &\frac{\partial \omega\pen^n}{\partial t} = -\frac{\gv{\nabla} \rho^n}{\rho^n}
    \times \left(\frac{\partial \gv{v}\pen^n}{\partial t} +
    \bv{\nabla} \cdot \left( \gv{v}\pen^n \otimes \gv{v}\pen^n \right) - \gv{b}^n \right)
    \label{update_baro} \\
    \textrm{Volumetric force term based vorticity update} ~~~ &\frac{\partial \omega\pen^n}{\partial t} = \gv{\nabla} \times \gv{b}^n
    \label{update_body} \\
    \textrm{Vorticity advection and remeshing} ~~~ &\frac{\partial \omega\pen^n}{\partial t} +
    \left( \gv{v}\pen^n \cdot \bv{\nabla} \right) \omega\pen^n = 0
    \label{vort_advect} \\
    \textrm{Vorticity propagation to next time step} ~~~ &\omega^{n + 1} = \omega\pen^{n + 1}
    \label{vort_reset} \\
    \textrm{Rigid body translational update} ~~~ &\gv{x}^{n+1}_{\mathrm{cmr}, j} = \gv{x}^{n}_{\mathrm{cmr}, j} +
    \gv{\dot{x}}_{\textrm{cmr}, j}^n \Delta t^n
    \label{cm_update} \\
    \textrm{Rigid body rotational update} ~~~ &\gv{\theta}^{n+1}_{r, j} = \gv{\theta}^{n}_{r, j} + \gv{\dot \theta}^n_{r, j} \Delta t^n
    \label{ang_update}
\end{align}
\end{algorithm}

\subsection{Projection and Brinkman penalization of rigid body motion}

For each rigid body $j$ in the simulation, we recover its level set per~\cref{chi_fix},
followed by projection of translational $\gv{\dot{x}}_{\textrm{cmr}, j}^n$
(\cref{lin_project}) and rotational $\gv{\dot \theta}_{r, j}^n$
(\cref{ang_project}) velocities. The volume integrals are carried out using the
mid-point rule with a discrete, mollified Heaviside integrand \( H_{\varepsilon} \)
as defined in \citet{gazzola2011simulations},
where the mollification length $\varepsilon = 2\sqrt{2}h$ is set
throughout the paper, $h$ being the grid spacing. Rigid components of motion
$\gv{\dot{x}}_{\textrm{cmr}, j}^n$ and $\gv{\dot \theta}_{r, j}^n$ so obtained are then
employed to determine the rigid velocity fields $\gv{v}_{r, j}$, which are fed to the
penalization operator (\cref{v_pen}), to finally recover the physically consistent flow field $\gv{v}\pen^n$.
The penalization operator for the velocity
field is formulated through a first order implicit Euler time discretization scheme
\cite{gazzola2011simulations}, to relax stability conditions related to the stiffness of
the penalization parameter $\lambda = 1e^4$ (throughout the paper). The additional vorticity
caused by penalization $\gv{\nabla} \times (\gv{v}\pen^n - \gv{v}^n)$ is added to the
unpenalized vorticity $\omega^n$ via \cref{vort_pen}. This approach avoids additional
diffusion of the \(\vort\) field as reported in~\citet{rasmussen2011multiresolution}.

\subsection{Inverse map advection}\label{sec:eta_advect}

After computing translational and rotational rigid body velocities and penalizing the flow
field accordingly, we consider the elastic bodies present in the domain. We start by
advecting the inverse map through \cref{xi_advect}. We do so by discretizing
$\gv{\nabla} \gv{\xi}$ using a WENO5 stencil \cite{liu1994weighted} and performing the temporal
integration using a SSP (Strong Stability Preserving) third order Runge-Kutta scheme
\cite{shu1989efficient}. The rationale behind this choice, as opposed to the use of
particles and remeshing, stems from two observations. First, particle advection and
moment conserving remeshing solve the conservative form of the advection
equation, which includes the additional term $\gv{\xi} ~(\gv{\nabla} \cdot \gv{v})$,
relative to \cref{xi_advect}. Although this extra term is zero for incompressible solids
and fluids, its numerical discretization leads to a high wave-number instability arising from the
solid--fluid interface. This is due to the localized and bounded
numerical incompressibility inconsistencies that stem from the blending between solid and
fluid phases (see \cref{app:mass_convg}). While modifications to the conservative form have been suggested
to mitigate the issue \cite{jain2019conservative}, these cannot be directly translated to
particle methods.
Second, as observed in \citet{hieber2008lagrangian}, advecting solid deformation maps using
particles does not relax the time step restriction dictated by the solid shear wave speed,
thus providing little incentive for particles over a convenient grid based non-oscillatory advection scheme.
Hence, our choice of WENO5 in combination with SSP-RK3. We also note that the advection of
the inverse map $\gv{\xi}$ (\cref{xi_advect}), unlike the advection of all other quantities
(\cref{vort_advect,cm_update,ang_update}), is executed early on
before the evaluation of Cauchy stress terms (\cref{update_stress}) and baroclinic terms
(\cref{update_baro}). This is because of the fundamental difference in the formulation
of flow--structure interaction in our algorithm, for rigid vs elastic bodies. For rigid bodies,
we purposefully advect all relevant quantities based on the previous time step's velocity field, to
leverage the mismatch in interface position for the recovery, through projection, of
the fluid forces acting upon the body. On the other hand, the elastic body--fluid interaction
necessitates the evaluation of explicit forces and torques at the latest solid
configuration (\cref{blend_stress}), which can only be computed by first advecting the
inverse map $\gv{\xi}$ using \cref{xi_advect}, to obtain the solid configuration  at the current time step.

\subsection{Level set recovery and reinitialization}

Using the advected \refmap we can reconstruct the deformed solid interface at
the next time step using~\cref{pin_phi}. However discretization errors corrupt
this reconstructed level set field \cite{sethian1999level} prompting the need to reinitialize it to restore the
signed-distance property of \cref{reinit_phi}. Here we utilize the second order accurate variant of
the fast marching method (FMM) described in \cite{sethian1999level} to reinitialize \( \phi_{e, i}\),
in a narrow band of 8 points on either side of the solid zone.

\subsection{\Refmap extrapolation}

As specified in~\cref{sec:invmap}, \( \gv{\xi}\) is only defined inside the solid
phase \( \Omega_{e}\). However, to numerically evaluate and eventually merge
stresses between solid and fluid, it is necessary to \emph{extend}
\( \gv{\xi}\) into the nearby fluid zone. This step, indicated in \cref{extrap_eta}, is
achieved by using the least-squares based extrapolation procedure reported by
\citet{jain2019conservative}.
In this work, we extrapolate information across 6 grid points, compatibly with the stencil
support of the spatial operators that will act on it.

\subsection{Stress evaluation and vorticity update}\label{sec:stressedout}

Here, we elaborate on the numerics involved in \cref{blend_stress,blend_rho,update_stress}.
First, the deformation gradient tensor $\bv{F}$ and the strain rate tensor $\bv{D}\dev$
are computed by taking derivatives of $\gv{\xi}$ and $\gv{v}$ respectively,
using second order centered finite differences.
Following the computation of $\bv{F}$ and $\bv{D}\dev$, we then compute the solid and
fluid stresses $\bv{\sigma}\dev_e$ and $\bv{\sigma}\dev_f$, using the solid and fluid
constitutive laws given in \cref{eqn:cons_fluid,eqn:cons_solid,eqn:solid_stress}.
A mollified Heaviside function as defined in \citet{gazzola2011simulations}
is then used to blend the solid
and fluid stresses (\cref{eqn:blend_stress}) and density (\cref{eqn:blend_rho}) to obtain
monolithic stress and density fields. We then compute the Cauchy stress contribution shown in
\cref{update_stress} in two steps. In the first step, we compute $\bv{\nabla} \cdot \bv{\sigma}\dev$
by taking derivatives of $\bv{\sigma}\dev$ using second order centered finite
differences.
In the second step, we compute $\bv{\nabla} \times (\bv{\nabla} \cdot \bv{\sigma}\dev)$ by
replacing the curl operator $\bv{\nabla} \times$ with its discrete second order centered
finite difference counterpart. The Cauchy stress contribution is then added to
the vorticity in a forward Euler step. With regards to this last step, we recommend, as
already noted in \cref{sec:eta_advect}, that the solid stress should be evaluated after the
\refmap advection step \cref{xi_advect}. We have observed this ordering to be robust and
numerically stable.

Finally, we bring attention to the specific domain of viscosity dominated problems, where
the bodies and the fluid have the same density ($\rho_{e, i} = \rho_{r, j} = \rho_f = \rho$) and
the same dynamic viscosity ($\mu_{e, i} = \mu_f = \mu$). In this case,
\cref{update_stress} simplifies to the following equation

\begin{equation}
    \label{eqn:per_diffusion}
    \frac{\partial \omega\pen^n}{\partial t} = \frac{1}{\rho} \left(\gv{\nabla}
    \times \gv{\nabla} \cdot \sum_i H(\phi_{e, i}^n) ~ \bv{\sigma}\dev_{e, i}^n \right) +
    \frac{\mu}{\rho} \bv{\nabla}^2 \omega\pen^n.
\end{equation}
As a consequence, in the case of a periodic domain, the diagonality of the $\bv{\nabla}^2$ operator in the
Fourier space (RHS of \cref{eqn:per_diffusion}) can be leveraged to achieve
unconditionally stable temporal integration of the viscous stress contribution to the
vorticity field, as detailed in \citet{kolomenskiy2009fourier}. Hence, the Fourier
condition (\cref{sec:timestep}) for the explicit update of viscous stresses
can be sidestepped to achieve faster time-to-solutions.

\subsection{Baroclinic and volumetric force terms}

We compute the baroclinic contribution of \cref{update_baro} as in
\citet{gazzola2011simulations}. We use a discrete second order centered
finite difference counterpart for the gradient operator \( \nabla \), while the temporal
derivative is approximated to first order using the differences
$\gv{v}_{\lambda}^{n} - \gv{v}_{\lambda}^{n-1}$. The baroclinic contribution
is then added to the vorticity in a forward Euler step. The same technique is
used to evolve the vorticity generated from the volumetric forcing terms
~\cref{update_body}. Lastly, in the case of a collision between two bodies, \cref{eqn:coll} is
used to compute collision forces, substituting the Dirac Delta function with its mollified
equivalent $\delta_{\varepsilon}$, as defined in \citet{jain2019conservative}.


\subsection{Vorticity advection and remeshing}

After updating the vorticity on the grid, we discretize it into particles of strength
\[ \Gamma_{p}=\sum_{i} \vort_{i} W\left(\frac{\gv{x}_{i}-\gv{x}_{p}}{h}\right) \]
where \( i \) symbolizes the grid index, \( p \) is the particle and \( W (\cdot)\) is
an interpolation kernel. In this work we used the fourth order \( M'_4 \) with
\( W(\gv{x}) = M'_4(x)M'_4(y)\) which conserves the first three moments
\cite{monaghan1985extrapolating}.
These particles are then advected using a third order Runge-Kutta scheme. The vorticity carried by
the particles is then remeshed at the grid nodes via the same interpolation
kernel and carried forth to the next time step (\cref{vort_reset}).

\subsection{Rigid body update}

Finally we evolve the position and orientation of all rigid bodies in the simulation
using the first order explicit Euler time integration scheme for reasons
detailed in \citet{gazzola2011simulations}.

\subsection{Restrictions on simulation time step}\label{sec:timestep}

We encounter four major time step restrictions in our algorithm due to the presence of
different time scales in the coupling problem. The first restriction is associated with
particle advection and remeshing (\cref{vort_advect}). This restriction is not dictated by the
usual CFL (Courant--Friedrich--Lewy) condition. Instead \( \Delta t \) is constrained  by the
amount of shear through the Lagrangian LCFL condition, which is independent of grid spacing \( h \)
\begin{equation}
    \label{LCFL}
    \Delta t_1 \leq \LCFL ~ || \omega ||_{\infty}^{-1}.
\end{equation}
Physically, this condition necessitates that particle remeshing kernels should always
overlap in space at all times. The independence from \( h \) indicates that our
particles based approach can take larger-than-CFL stable time steps, reducing time-to-solution.
A second restriction stems from the need to resolve shear waves inside elastic solids.
This is a CFL-like restriction dependent on the shear wave speed $c_{sh}$
\begin{equation}
    \label{shear_CFL}
    \Delta t_2 \leq  h ~ \CFL  ~ c_{sh}^{-1} = h ~\CFL  ~ \sqrt{\rho_s / G}.
\end{equation}
Here, $\rho_s$ and $G$ correspond to the solid density and shear modulus, respectively.
Another related restriction stems from the advection of the \refmap $\gv{\xi}$ inside the elastic solid
\begin{equation}
    \label{xi_CFL}
    \Delta t_3 \leq \CFL ~ (h) ~ ||\gv{v}_{e}||_{\infty}^{-1}
\end{equation}
where $\gv{v}_e$ refers to the velocity field inside the elastic solid. We observe that in
most cases $\Delta t_3 > \Delta t_2$, rendering the condition on shear waves more
stringent. Finally, the fourth restriction
is the Fourier condition that ensures stability with regards to explicit time discretization
of the viscous stresses inside both the solid and fluid
\begin{equation}
    \label{diff_limit}
    \Delta t_4 \leq k ~ (h)^2 / 4 \max(\nu_{f}, \nu_{e})
\end{equation}
where $k$ is a constant usually set to be \( \leq 1\). Here we set $k = 0.9$  throughout.
We note that in purely periodic
domains with uniform viscosity, we can utilize an implicit discretization of the viscous terms
(\cref{sec:stressedout}), effectively side-stepping this restriction. Combining
\cref{LCFL,shear_CFL,xi_CFL,diff_limit}, we obtain the final criterion to adapt the time
step during simulation
\begin{equation}
    \Delta t = \min{\left[ \Delta t_1, \Delta t_2, \Delta t_3, \Delta t_4 \right] }.
\end{equation}

Following a detailed description of our algorithm, we now investigate the accuracy and
convergence properties of our algorithm, via extensive validation across analytical and numerical benchmarks.

\section{Validation benchmarks}\label{sec:bmks}

We now proceed to validate the proposed method across several benchmark cases.
These involve a pure solid system, forced oscillations in parallel layers of fluid and
solid, fluid induced shape oscillations of a visco-hyperelastic cylinder, and collision
between two hyperelastic cylinders surrounded by fluid. For all these cases,
the dimensional parameters are specified in SI units, unless otherwise noted.
Additionally, they all utilize a square computational domain of unit dimension \( [0, 1]^2 \).
For each case, we conduct a convergence analysis by reporting discrete
\Ltwo and \Linf error norms of relevant physical quantities as a function of spatial and
temporal discretization. We use the following definition of discrete norms

\begin{equation}
    \Ltwo(e) := \norm{e}_{2} = \sqrt{h^2 \sum_{i} \left\lvert e_{i} \right\rvert^2 }; ~~~
    \Linf(e) := \norm{e}_{\infty} = \max_{i} \left\lvert e_{i} \right\rvert; ~~~
    e := p - p_{\textrm{ref}}
\label{eq:bmkint}
\end{equation}
where \(e\) denotes the error, \(p\) is a physical quantity obtained from our
method, \( p_{\textrm{ref}}\) is the reference solution, \( h\) denotes
grid spacing and \( i \) denotes the grid point index, unless otherwise noted.

Depending on the specific problem, the dynamics at play may be governed by one or more key
dimensionless numbers. We list them here, together with their physical interpretation

\begin{equation}
        \Rey := ~~~~\frac{\rho_f V L}{\mu_f} \sim\frac{\textrm{inertial
        forces}}{\textrm{viscous forces}}; ~~~~
        \Ca := ~~~~\frac{\rho_e V^2}{G} \sim\frac{\textrm{inertial
        forces}}{\textrm{elastic forces}}; ~~~~
        \Er := ~~~~\frac{\mu_f V}{G L} \sim\frac{\textrm{viscous forces}}{\textrm{elastic forces}}
\end{equation}
where \Rey, \Ca, \Er, $V$, $L$, $\mu_f$, $\rho_f$, $\rho_e$ and $G$ correspond to the
Reynolds number, Cauchy number \cite{massey1998mechanics}, Ericksen number \cite{larson1993ericksen},
velocity scale, length scale, fluid viscosity, fluid density, elastic solid density and
shear modulus of the solid, respectively.

\subsection{Pure solid system}\label{sec:bmk1}

\begin{figure}[!ht]
    \centerline{\includegraphics[width=\textwidth]{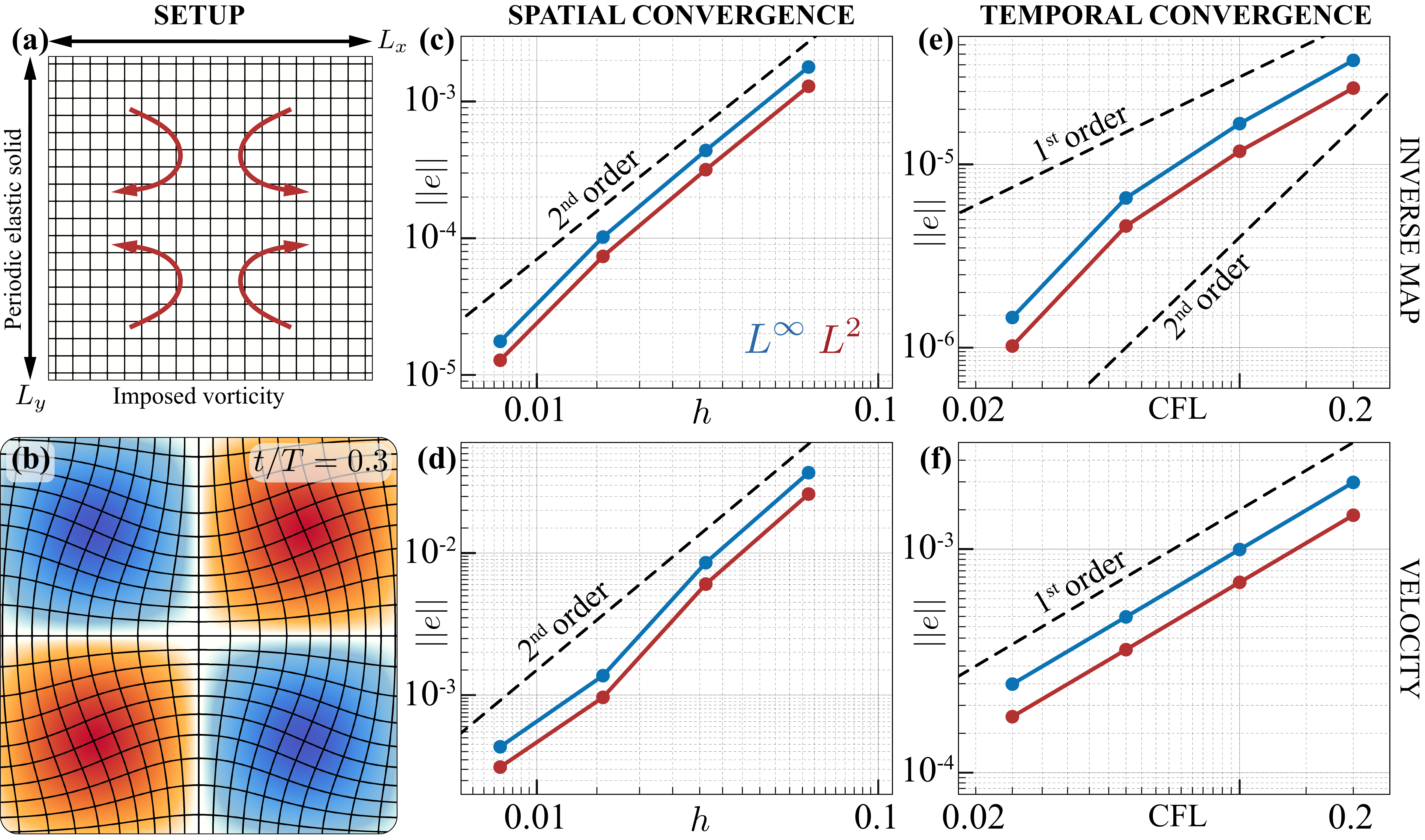}}
    \caption{Pure solid system. \capsub{a} Case setup. The neo-Hookean solid used
        is purely hyperelastic (\(\mu_{e} = 0\)), has $\rho_e = 1$ and has shear modulus \( c_1 =
        0.5\) i.e. \( G = 2c_1 = 1.0\). The parameters for the imposed vorticity
        are $\omega_0 = 0.05$, $T = 1$ and $L_x = L_y = L = 1$. The key non-dimensional
        parameter for this benchmark is $\Ca = \rho_e \omega_0^2 L^2 / G = 5 \times
        10^{-4}$.
        \capsub{b} Deformed solid system with inverse map (black lines) and imposed
        vorticity field (orange/blue represent positive/negative vorticity)
        contours at $t/T = 0.3$. Spatial convergence: $L_{\infty}$ (blue) and $L_2$ (red)
        norms of the error are plotted against grid spacing $h$, for \capsub{c} inverse map and
        \capsub{d} velocity, respectively. Temporal convergence: $L_{\infty}$ (blue) and $L_2$ (red)
        norms of the error are plotted against $\CFL$, for \capsub{e} inverse map and
        \capsub{f} velocity, respectively.}
    \label{fig:bmk1}
\end{figure}

We first test our method for the case of a pure solid system, previously reported by
\citet{zhao2008fixed}. This case utilizes the components of the algorithm only
pertaining to the solid phase---Poisson solve, solid stress update, diffusion and
advection---the other components will be analyzed in the subsequent benchmarks.
The neo-Hookean solid, shown in \cref{fig:bmk1}a, is initialized to be stress free and
entails periodic boundaries. Following the method of manufactured solutions
\cite{roache2002code}, we first derive a semi-analytical
reference solution against which we validate our solver. We start by computing the
semi-analytical inverse map $\gv{\xi}_{\textrm{sa}}$ resulting from the advection of
$\gv{\xi}^0 = (X , Y)$ through the imposed velocity field

\begin{equation}
    \begin{aligned}
        \gv{v}_{\textrm{ref, x}}(t) &= \frac{L_x^2 L_y}{2 \pi (L_x^2 + L_y^2)} \omega_{0} \sin (2 \pi t / T) \sin
        (2 \pi x / L_{x}) \cos (2 \pi y / L_{y}) \\
        \gv{v}_{\textrm{ref, y}}(t) &= \frac{-L_x L_y^2}{2 \pi (L_x^2 + L_y^2)} \omega_{0} \sin (2 \pi t / T) \cos
        (2 \pi x / L_{x}) \sin (2 \pi y / L_{y})
    \end{aligned}
  \label{eq:bmk1_0}
\end{equation}
which corresponds to the vorticity field

\begin{equation}
    \omega_{\textrm{ref}}(t)= \omega_{0} \sin (2 \pi t / T) \sin (2 \pi x / L_{x}) \sin (2 \pi y / L_{y})
  \label{eq:bmk1_1}
\end{equation}
where $\omega_0$ is a constant, $L_x$ and $L_y$ are the dimensions of the computational
domain,
and $T$ is the time period of the imposed motion. Details relative to these quantities and the
computational setup can be found in the caption of
\cref{fig:bmk1}. We then calculate the external body force $\gv{b}_{\mathrm{ext}}(\gv{x}, t)$
that needs to be applied to the solid at rest $\gv{\xi}^0$ to reproduce the above motion

\begin{equation}
    \gv{b}_{\mathrm{ext}} = \frac{\partial \omega_{\textrm{ref}}}{\partial t} + \gv{v}_{\textrm{ref}} \cdot
    \gv{\nabla} \omega_{\textrm{ref}} -\gv{\nabla} \cdot \bv{\tau}_{\mathrm{elas}}\left(\gv{\xi}_{\textrm{sa}}\right)
  \label{eq:bmk1_2}
\end{equation}
where the dependence of solid stress \( \bv{\tau}_{\mathrm{elas}} \) on the previously
computed \refmap \( \gv{\xi}_{\textrm{sa}} \) is made explicit. All operators are
either analytical or discretized as in \cref{sec:num}.
We then perform a separate simulation using our numerical method, in which we apply this
force $\gv{b}_{\mathrm{ext}}(\gv{x}, t)$ to the solid at rest and record the output
numerical velocity \(\gv{v} \) and inverse map $\gv{\xi}$ at a prescribed point of time.
The convergence order for both the inverse map $\gv{\xi}$ and velocity field $\gv{v}$
is finally determined by computing the \Ltwo and \Linf norms (\cref{eq:bmkint}) relative
to the semi-analytical inverse map \( \gv{\xi}_{\textrm{sa}} \)
(at the finest resolution $512 \times 512$, $\CFL, \LCFL = 0.1$)
and the analytical velocity field $\gv{v}_{\textrm{ref}}$. For spatial
convergence, we fix $\Delta t$ based on $\CFL = \LCFL = 0.05$ at grid resolution $128
\times 128$, and vary the spatial resolution between $16 \times 16$ and $128 \times
128$. For temporal convergence instead, we set the spatial
resolution to $256 \times 256$ and vary the $\CFL = \LCFL$ between 0.025 and 0.2.
As seen in \cref{fig:bmk1}c,d the method presents second order spatial convergence
for both the inverse map and velocity field, which is in agreement with our spatial
discretization of the operators. Temporal convergence (\cref{fig:bmk1}e,f) is instead
found to be between first and
second order (least squares fit of 1.5) for the inverse map and first order for the velocity
field, as expected from the Godunov splitting adopted in the timestepping algorithm
(\cref{update_stress,update_baro,update_body,vort_advect}).

\subsection{Oscillatory response in parallel layers of fluid and solid}\label{sec:bmk2}

\begin{figure}[!ht]
    \centerline{\includegraphics[width=\textwidth]{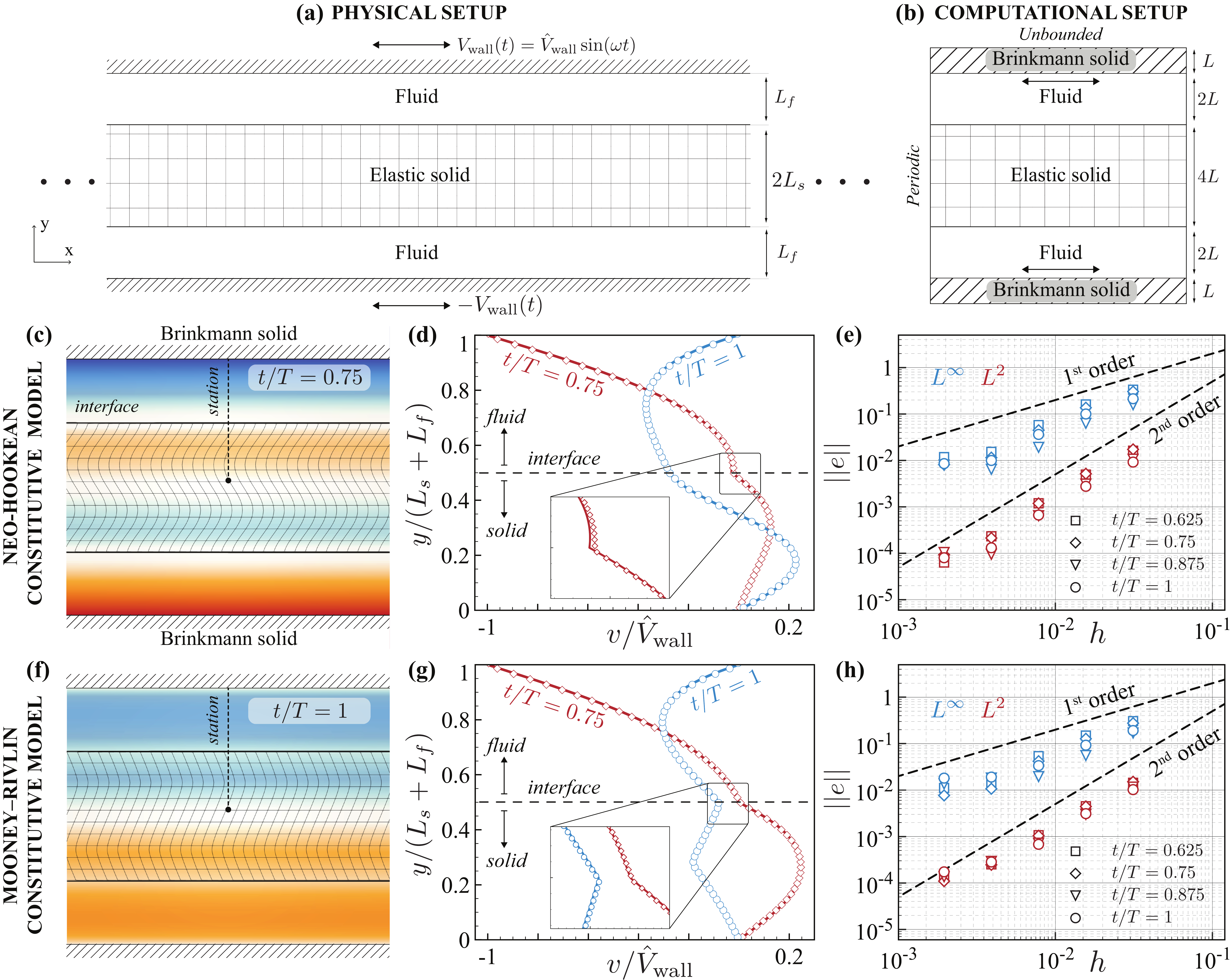}}
    \caption{
    Oscillatory response in parallel solid--fluid layers. \capsub{a} Physical
    setup of the parallel solid--fluid layers, with the walls moving
    sinusoidally in opposite directions with an imposed velocity \(
    V_{\textrm{wall}}(t) := \pm \hat{V}_{\textrm{wall}}\sin(\omega t) \). \capsub{b}
    Computational setup indicating the boundary conditions and the use of a
    Brinkman solid to model the walls. Here, we
    set the thickness of the elastic slab to \( 2L_{s} = 4L\), each fluid layer
    to \( L_{f} = 2L \) and each Brinkman solid layer to \( L \), with \( L =
    0.1\). The system is hence symmetric about the mid-plane. The kinematic
    parameters used in the simulation are \( \omega = \pi, T = 2\pi/\omega = 2,
    \hat{V}_{\textrm{wall}} = 0.4 \), leading to a shear rate of \( \dot{\gamma}
    = {\hat{V}_{\textrm{wall}}} / {(L_{s} + L_{f})} = 1 \). The dynamic
    parameters corresponding to the fluid phase are \( \rho_{f} = 1, \mu_{f} =
    0.02\). The neo-Hookean visco-elastic solid used in \capsub{c-e}
    is density matched (\( \rho_{f} = \rho_{e} \)), has dynamic viscosity
    \(\mu_{e} = 0.1\mu_{f} \) and has shear modulus \( c_1 = 0.01\)
    i.e. \( G = 2c_1 = 0.02\). For the generalized Mooney--Rivlin
    material used in \capsub{f-h}, we retain the parameters above and
    additionally set \( c_2 = 0, c_3 = 4 c_1 \). The meanings of these coefficients are
    detailed in \cite{bower2009applied}. The simulations are run
    until \(t / T = 10 \), and physical quantities are sampled
    within the last cycle. The key non-dimensional dynamic parameters for this
    benchmark are \( \Rey = {\rho_{f} \dot{\gamma} L^2_{f}} / {\mu_{f}} = 2 , \Er
    = {\mu_{f} \hat{V}_{\textrm{wall}} } / {2 G L_{s}} = 1\). The computational
    parameters are set to \( \LCFL = 0.05, \CFL = 0.1\).
    \capsub{c, d, e} Results for the solid with neo-Hookean constitutive law.
    \capsub{c} Velocity field (orange/blue represent positive/negative velocity)
    and \refmap contours within the domain, with the interface marked (black, thick solid) for visual clarity.
    Upon plotting the velocity profiles at the highlighted station (black,
    dashed) in the center of the domain, good agreement with analytical results is
    observed across all times as shown in \capsub{d}, which plots the non-dimensional
    station position versus \(x\)-velocity. The inset shows the concentration of errors
    near the diffuse interface. For reference, numerical results are plotted with scatter points whereas
    analytical results are plotted with a solid line.
    Tracking these results with changing
    resolution results in the convergence plot shown in \capsub{e} where \Linf (blue) and \Ltwo (red)
    norms of the error are plotted against grid spacing $h$ at different \( t /
    T\). Trends indicate a first to second order convergence as expected. \capsub{f, g, h}
    Results for the solid with generalized Mooney--Rivlin constitutive law are
    found to be consistent with the above trends.
    }
\label{fig:bmk2}
\end{figure}

Having tested the ability of our method to capture purely elastic responses of
the solid media, we now proceed to validate the interfacial coupling between solid
and fluid phases. We adopt the benchmark setup shown in
\cref{fig:bmk2}a, first proposed by \citet{sugiyama2011full}. Here an elastic solid layer
is sandwiched between two fluid layers, in turn confined by two long planar
walls, whose horizontal oscillations drive a characteristic system response. Indeed, this
setting admits a time periodic, one-dimensional analytical solution, which we generalised
from \citet{sugiyama2011full} to include visco-hyperelastic, density mismatched solid
\cite{parthasarathy2020simple}.
Overall, this problem entails
multiple interfaces, phases and boundary conditions interacting dynamically, and serves as
a challenging benchmark to validate the long time behaviour, stability and accuracy of our
solver.

We computationally realize this setup as shown in \cref{fig:bmk2}b. Instead of modelling
the walls as a kinematic condition at the boundaries of the fluid phase, we actually
represent the walls within the computational domain as Brinkman solids.
This choice enables us to test rigid solid,
elastic solid and fluid coupling in the same simulation while demonstrating the
flexibility of our method. Then, periodic and unbounded boundary conditions are
imposed in the \( x \) and \( y \) directions respectively \cite{chatelain2010fourier}.
We investigate two separate cases in which the density matched visco-elastic solid is
either made of a neo-Hookean material or a generalized
Mooney-Rivlin material \cite{bower2009applied}. The system starts from rest in a
stress free state and the simulation is run well beyond the initial transient phase,
resulting in periodic dynamics. Details can be found in the figure caption.

\Cref{fig:bmk2}c showcases the numerical \(x\) velocity field and the \refmap
contours at the time of maximal deformation (\(t / T = 0.75\)) for the
neo-Hookean case. We plot the corresponding non-dimensional \(x\) velocities at the marked
line station for \(t / T = 0.75\) and \(t / T = 1\) in \cref{fig:bmk2}d, onto which the
analytical solution is overlaid. As it can be seen our simulations compare well with the
benchmark, with the maximum deviation occurring at the solid--fluid interface. This is
expected given that our approach involves a diffuse interface. We then plot the
\Ltwo and \Linf norms of the error \(e\) defined as \(e = \left(
\gv{v}_{\textrm{sim}} - \gv{v}_{\textrm{analytical}} \right) \cdot \hat{i}\)
at different time instances, for spatial resolutions between \(32 \times 32\)
and \(512 \times 512\) (\cref{fig:bmk2}e). The \Ltwo convergence is approximately
second order ($\Ltwo = 1.8$), while for \Linf it is closer to first order ($\Linf = 1.3$).
This is because of the localised errors at the interface, where a $C^1$ discontinuity of
the physical solution is observed.

\Cref{fig:bmk2}f-h refer to the generalized Mooney--Rivlin case. The effect of
solid non-linearity can be seen from the \refmap
contours (\cref{fig:bmk2}f) and corresponding velocities within the solid phase in
\cref{fig:bmk2}g, and manifest as a sharp ``bend'' in the solid midplane at \(y / \left(L_{f} + L_{s} \right) = 0.25\).
Once again, numerical and analytical results are in agreement. Convergence of errors
with spatial resolution (\cref{fig:bmk2}h) shows trends similar
to the case with the neo-Hookean constitutive model (1.5 for \Ltwo and 0.9 for \(\Linf\)).

The results of this section indicate the ability of our approach to successfully capture
fluid--elastic solid and fluid--rigid solid interactions that are themselves coupled.
Additionally, in \cref{app:rho_mismatch} we tested the effect of solid density mismatch
and observed consistent convergence and accuracy properties.

\subsection{Fluid induced shape oscillations of a visco-hyperelastic cylinder}\label{sec:bmk3}

\begin{figure}[!ht]
    \centerline{\includegraphics[width=\textwidth]{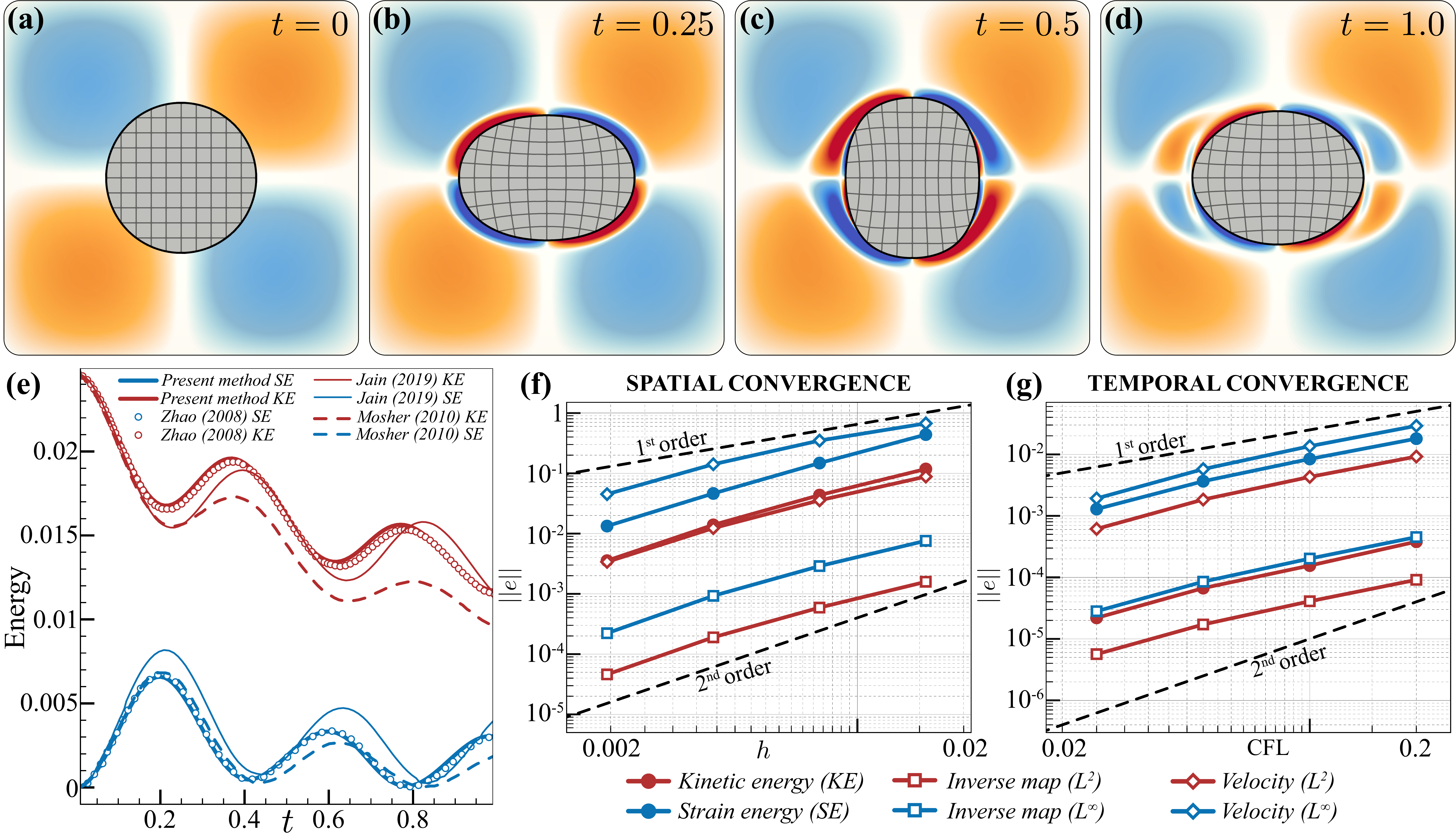}}
    \caption{Fluid induced shape oscillations of a visco-hyperelastic cylinder.
        \capsub{a} Case setup. The dynamic parameters corresponding to the
        fluid phase are \(\rho_{f} = 1\), \(\mu_{f} = 10^{-3}\). The neo-Hookean
        visco-hyperelastic cylinder, placed at (0.5, 0.5), has radius $r = 0.2$, is density matched
        ($\rho_s = \rho_f$), has dynamic viscosity $\mu_e = \mu_f$, and
        has shear modulus \( c_1 = 0.5\) i.e. \( G = 2c_1 = 1.0\).
        The parameters corresponding to the streamfunction for the imposed vorticity
        are $\psi_0 = 0.05$, and $L_x = L_y = L = 1$. The key non-dimensional parameters for
        this benchmark are $\Rey = \rho_f \psi_0 r / \mu_f L = 10$ and $\Ca = \rho_e
        \psi_0^2 / G L^2 = 5 \times 10^{-4}$. \capsub{b-d} Temporal variation of inverse
        map (black lines) and vorticity field (orange/blue represent positive/negative vorticity)
        contours, showing the dynamic response of the
        cylinder to the initial imposed vorticity. \capsub{e} Comparison of temporal variation
        of the kinetic energy $\mathrm{KE}$ and strain energy $\mathrm{SE}$
        with previous studies \cite{jain2019conservative, zhao2008fixed,
        robinson2011symmetric}.
        \capsub{f} Spatial convergence: $L_{\infty}$ (blue) and $L_2$ (red) norms of
        the error are plotted against grid spacing $h$, for the energies, inverse map
        and velocity.
        \capsub{g} Temporal convergence: $L_{\infty}$ (blue) and $L_2$ (red)
        norms of the error are plotted against $\CFL$, for the energies, inverse map and
        velocity.}
\label{fig:bmk3}
\end{figure}

We now test the capability of capturing dynamics related to time dependent geometrical
variations of a two-dimensional fluid--solid interface. To do so we adopt the benchmark setup
of a neutrally buoyant freely oscillating cylinder immersed in a fluid, first
reported by \citet{zhao2008fixed}. \Cref{fig:bmk3}a highlights the initial physical
setup---a stress free cylinder surrounded by fluid is placed at the center of the domain with
periodic boundaries.
We then deform the solid through an initial imposed Taylor--Green vorticity field,
corresponding to the streamfunction profile
\begin{equation}
    \psi =\psi_{0} \sin (2 \pi x / L_{x}) \sin (2 \pi y / L_{y})
  \label{eq:bmk3_1}
\end{equation}
where $\psi_0$ is a constant and $L_x$, $L_y$ are the dimensions of the computational domain
(details in \cref{fig:bmk3}).

\Cref{fig:bmk3}b-d showcase the temporal evolution of inverse map, vorticity contours
and the observed dynamics of the cylinder, which resembles a damped oscillator. Deformed initially
by the imposed vorticity, the cylinder retracts due to its elastic response.
This sets up oscillations, which slowly decay over time as the solid dissipates its elastic
potential energy due to viscous effects.
We then track the temporal variation of the kinetic energy of the system and strain energy
of the solid, and compare with previous calculations based on
finite elements \cite{zhao2008fixed, robinson2011symmetric} and finite volumes
\cite{jain2019conservative}. System kinetic energy $\mathrm{KE}$ and solid strain energy
$\mathrm{SE}$ are defined and discretized as follows
\begin{equation}
        \mathrm{KE} = \int_{\Sigma} \frac{1}{2} \lvert{\gv{v}}\rvert^2 \ d\gv{x} \approx
        \frac{h^2}{2} \sum_{i} \lvert{\gv{v}_i}\rvert^2; ~~~~
        \mathrm{SE} = \int_{\Omega_e} c_1 ({tr}(\bv{F}^{T} \bv{F}) - 2) \
        d\gv{x} \approx c_1 h^2 \sum_{i} \textrm{sgn}(\phi_i) ({tr}(\bv{F}_i^{T} \bv{F}_i) - 2)
  \label{eqn:bmk3_2}
\end{equation}
As can be seen in \cref{fig:bmk3}e, our results are found to be consistent with the other
methods, and in particular in close agreement with \citet{zhao2008fixed}.

We then present spatial and temporal convergence of energies,
inverse map $\gv{\xi}$ and velocity field $\gv{v}$ at $t = 0.25$, by computing the $L_2$ and $L_{\infty}$
norms of the error field, with respect to the best resolved case. For spatial
convergence, we fix $\Delta t$ based on $\CFL = \LCFL = 0.2$ for the grid resolution $1024
\times 1024$, and vary the spatial resolution between $32 \times 32$ and $512 \times
512$ (with $1024 \times 1024$ as the best resolved case). For temporal convergence
instead, we set the spatial resolution to $256 \times 256$ and vary the $\CFL = \LCFL$
between 0.2 and 0.025 (with $\CFL = \LCFL = 0.0125$ as the best resolved case).
As seen from \cref{fig:bmk3}f, the method presents spatial convergence between first and
second order ($\Ltwo = \Linf = 1.5$) for energies and inverse map. The convergence
order for the velocity field was found to be first order for $\Linf$ and 1.4 for for $\Ltwo$.
As shown in \cref{fig:bmk3}g, the temporal convergence order was found to be between first
and second order ($\Ltwo = \Linf = 1.3$), for all concerned quantities.

\subsection{Collision between two hyperelastic cylinders immersed in a fluid}\label{sec:bmk4}

\begin{figure}[!ht]
    \centerline{\includegraphics[width=\textwidth]{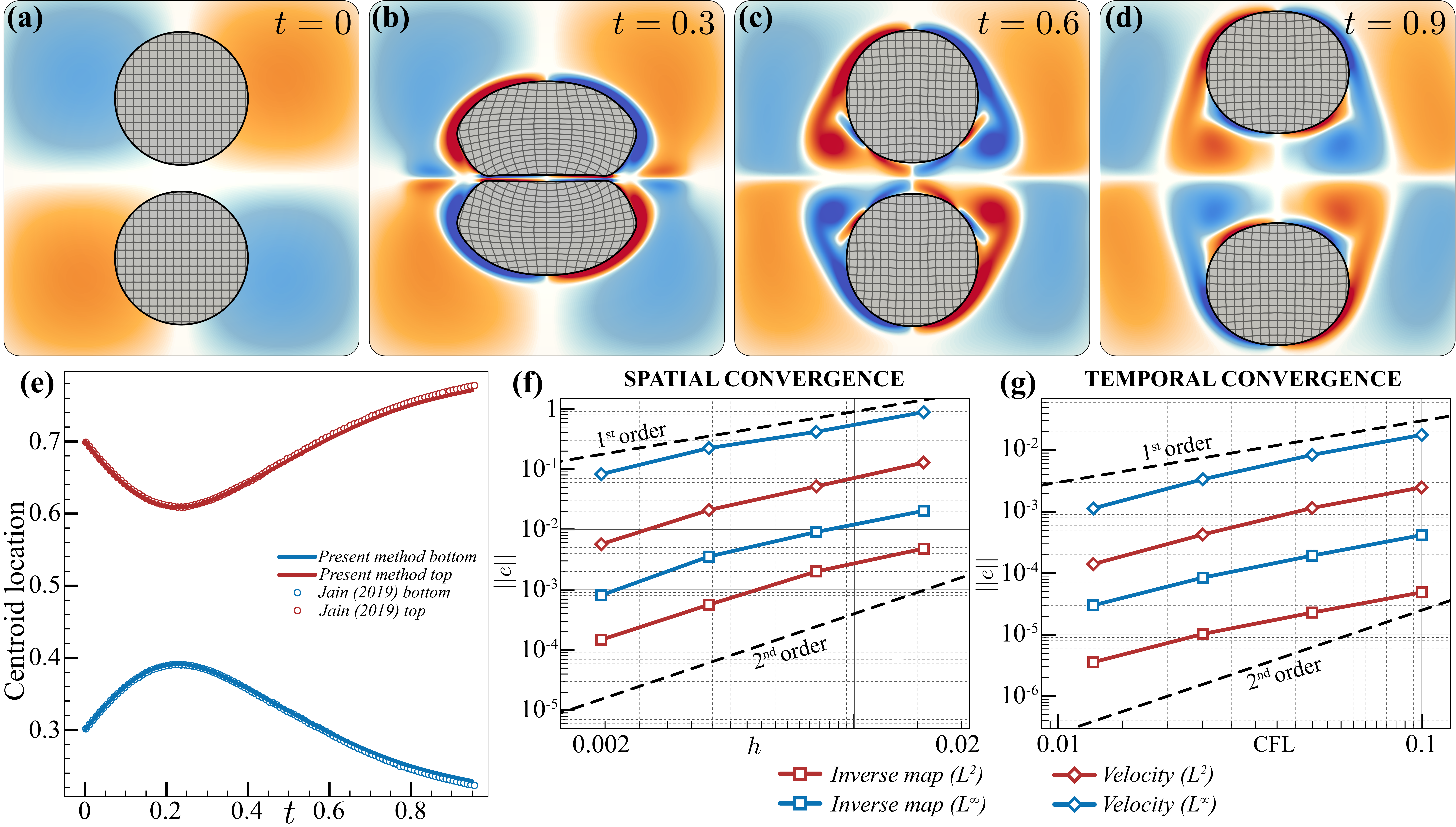}}
    \caption{Collision between two hyperelastic cylinders immersed in a fluid.
        \capsub{a} Case setup. The dynamic parameters corresponding to the
        fluid phase are \(\rho_{f} = 1\), \(\mu_{f} = 10^{-2} / 2 \pi\). The neo-Hookean
        hyperelastic disks, located at (0.5, 0.3) and (0.5, 0.7),
        have radii $r = 1 / 6$, are density matched
        ($\rho_e = \rho_f$), have no internal dissipation ($\mu_e = 0$), and
        have shear modulus \( c_1 = 1.0\) i.e. \( G = 2c_1 = 2.0\).
        The parameters corresponding to the streamfunction for the imposed vorticity
        are $\psi_0 = 1 / 2 \pi$, and $L_x = L_y = L = 1$. The key non-dimensional parameters for
        this benchmark are $\Rey = \rho_f \psi_0 r / \mu_f L = 16.67$ and $\Ca = \rho_e
        \psi_0^2 / G L^2 = 0.0127$. \capsub{b-d} Temporal variation of inverse
        map (black lines) and vorticity field (orange/blue represent positive/negative
        vorticity) contours, showing the dynamic response of the
        cylinders to the initial imposed vorticity. \capsub{e} Comparison of temporal variation
        of the centroids of the cylinders with previous studies \cite{jain2019conservative}.
        \capsub{f} Spatial convergence: $L_{\infty}$ (blue) and $L_2$ (red) norms of
        the error are plotted against grid spacing $h$, for the inverse map and velocity.
        \capsub{g} Temporal convergence: $L_{\infty}$ (blue) and $L_2$ (red)
        norms of the error are plotted against $\CFL$, for the inverse map and velocity.}
        \label{fig:bmk4}
\end{figure}

Following successful validation of our method for a single elastic body--fluid
interaction, we now demonstrate the ability of our solver to capture interactions between
multiple elastic bodies immersed in a fluid. Accordingly, we reproduce the
case of collision between two hyperelastic neo-Hookean cylinders in a fluid, first reported by
\citet{jain2019conservative}. Additionally, this case also highlights the capability of
our solver to simulate purely hyperelastic solids in a numerically stable fashion without the need for
internal viscous dissipation $\mu_s$. \Cref{fig:bmk4}a presents the initial physical setup with two
stress free neutrally buoyant circular cylinders immersed in a fluid, occupying a square domain
with periodic boundaries. The system then evolves due to an initial imposed
Taylor--Green vorticity field, corresponding to the streamfunction described in
\cref{eq:bmk3_1}. Computational setup details can be found in \cref{fig:bmk4}.

\Cref{fig:bmk4}b-d showcase the temporal dynamics of the two cylinders,
along with the inverse map and vorticity contours. The imposed
vorticity causes the two cylinders to collide, to then rebound due to both contact
forces and the internal stresses
generated as a result of the deformation. We validate our solver by comparing the
temporal variation of the centroids of both cylinders, against previous results
\cite{jain2019conservative}. As seen in \cref{fig:bmk4}e, our results show close agreement with
the benchmark \cite{jain2019conservative}.

We then present the spatial and temporal convergence of the
inverse map $\gv{\xi}$ and velocity field $\gv{v}$ at $t = 0.3$,
with respect to the best resolved case. For spatial
convergence, we fix $\CFL = \LCFL = 0.1$ and vary the spatial resolution
between $64 \times 64$ and $512 \times 512$ (with $1024 \times 1024$ as the best
resolved case). For temporal convergence instead, we set the spatial resolution to
$256 \times 256$ and vary the $\CFL = \LCFL$
between 0.1 and 0.0125 (with $\CFL = \LCFL = 0.00625$ as the best resolved case).
As seen from \cref{fig:bmk4}f, the method presents spatial convergence
between first and second order for inverse map (1.7 for \Ltwo and 1.5 for \Linf) and
velocity field (1.5 for \Ltwo and 1.1 for \Linf).
As shown in \cref{fig:bmk4}g, the temporal convergence order was found to be between first and
second order ($\Ltwo = \Linf = 1.3$) for inverse map and velocity field.
Additionally, in \cref{app:mass_convg} we report the convergence of
incompressibility errors in the solid, again found to be consistent with the above rates.

Overall the results of this section validate our algorithm against an extensive range of
benchmarks, showing the accuracy and robustness of our numerical scheme and its
implementation. These results are complemented by a detailed convergence analysis
which is found to be consistent with the employed discrete operators and across physical
scenarios. Critically, we demonstrated how our formulation naturally allows for the
seamless inclusion of a variety of physical phenomena within a consistent framework,
preserving stability, accuracy and convergence properties, thus enhancing usability and
utility. In the next section we expand on this, further illustrating the wide scope of our
solver in a range of multi-physics, complex problems.

\section{Numerical results: multi-physics illustrations}\label{sec:ills}

Next, we highlight the versatility of our solver by demonstrating a range
of potential applications. These include elasticity-induced viscous streaming phenomenon,
dynamic collision response of a ball falling under gravity on a soft trampoline,
dynamic and heat transfer characterization of an elastic flag flapping in the wake of a
hot cylinder and interaction between multiple, activated, self-propelling soft swimmers.

\subsection{Elasticity-induced viscous streaming}\label{sec:visc_stream}

Here we demonstrate the ability of our solver to successfully capture second order flow
physics effects and rectification phenomena, through the example of viscous streaming.
Viscous streaming refers to the time-averaged steady flow that arises when an immersed body
of characteristic length scale $a$ undergoes small-amplitude oscillations (compared to $a$)
in a viscous fluid. This phenomenon has found application in modern inertial
microfluidics, as an efficient, controllable mechanism for particle manipulation and sorting
\cite{lutz2005microscopic, marmottant2004bubble, liu2003hybridization}.
Viscous streaming has been well explored and characterized theoretically, experimentally
and computationally for rigid shapes of constant curvature such as cylinders or spheres
 \cite{lutz2005microscopic, stuart1966double, bertelsen1973nonlinear,
parthasarathy2018viscous, parthasarathy2019streaming, Riley:1966,Kotas:2007}, and more
recently in settings involving complex rigid geometries of multiple curvatures
\cite{parthasarathy2019streaming, bhosale2019, bhosale_parthasarathy_gazzola_2020}. Yet,
little is known regarding the streaming response to elastic body oscillations,
a potentially important aspect in scenarios involving biological materials
\cite{nagano2010full, wang1992nonlinear}. Motivated by this, we first attempt to
numerically recover the classic 2D rigid cylinder solution, to then explore the effect of
elasticity in the steady flow response.

\begin{figure}[!ht]
    \centerline{\includegraphics[width=\textwidth]{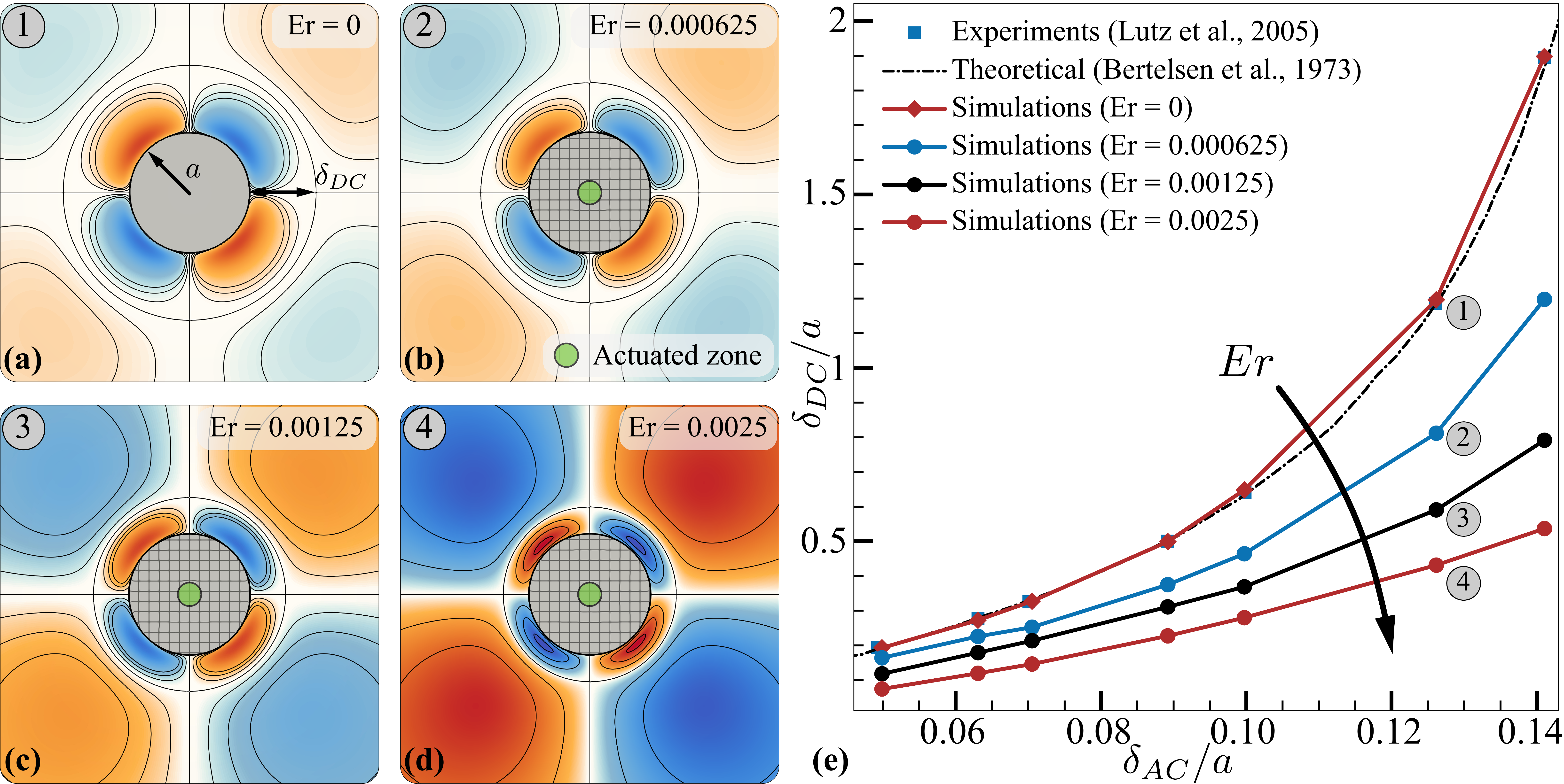}}
    \caption{Elasticity-induced viscous streaming. Case setup for \capsub{a} rigid
        cylinder and \capsub{b} visco-hyperelastic cylinder. Both the cylinders are
        density matched with the fluid ($\rho_e = \rho_r = \rho_f = 1$), have radius $a
        = 0.125$, and are placed at (0.5, 0.5) in the computational domain. The parametric
        values for the imposed motion $x(t) = x(0) + \epsilon a \sin(\omega t)$ with
        characteristic velocity $V_0 = \epsilon a \omega$ are $x(0)
        = 0.5$, $\epsilon = 0.1$ and $\omega = 32 \pi$. The rigid cylinder is modeled as
        a Brinkman solid while the visco-hyperelastic cylinder is
        modeled as a neo-Hookean solid with shear modulus $G$ ($c_1 = 2G$) and internal
        dissipation $\mu_s = \mu_f$. The actuation zone for the elastic solid is a
        cylinder with radius $r = 0.2 a$ and we did not see appreciable changes when varying this radius \( r \). Parametric values of $\mu_f$ and $G$ are
        determined based on the following key non-dimensional parameters:
        $\delta_{AC} / a = (\mu_f/ \rho_f \omega)^{1/2} / a$ and $Er = \mu_f V_0 / G a$.
        The computational parameters are set to $\LCFL = \CFL = 0.1$ with a
        spatial resolution of $1024 \times 1024$.
        Time averaged streamline patterns (blue/orange represent clockwise/anti-clockwise
        rotating regions) depicting streaming response at
        $R_s = 0.63$ ($\delta_{AC} / a = 0.126$) with increasing values of $Er$:
        \capsub{a} rigid body ($Er = 0$), \capsub{b} $Er = 0.000625$, \capsub{c} $Er =
        0.00125$ and \capsub{d} $Er = 0.0025$. The upper bound of $Er$ is chosen such that
        a finite thickness DC layer is observed.
        \capsub{e} Comparison of normalized DC boundary layer thickness $\delta_{DC} / a$
        vs. normalized AC boundary layer thickness $\delta_{AC} / a$ of our simulations
        (red diamonds) against experiments (blue squares \cite{lutz2005microscopic}) and
        theory (black dotted \cite{bertelsen1973nonlinear}), along with variation of the
        boundary layer scaling for different $Er$ values (blue, black and red circles).
        The shear modulus $G$ falls in the range $[15, 500]$ with variations in $\delta_{DC} / a$
        and $\Er$.
    }
    \label{fig:visc_stream}
\end{figure}

\Cref{fig:visc_stream}a,b highlight the physical setup---a circular
rigid or visco-hyperelastic cylinder of radius $a$ is placed at the centre of
a square domain with unbounded boundary conditions, under quiescent flow
conditions. We then impose a small amplitude oscillatory motion
$x(t) = x(0) + \epsilon a \sin(\omega t)$ with characteristic velocity $V_0 = \epsilon a
\omega$, where $\epsilon$ and $\omega$ are the non-dimensional amplitude
and angular frequency, respectively. In the rigid body limit, the cylinder is
formulated as a Brinkman solid and the entire body is actuated with the above motion.
For the visco-hyperelastic cylinder instead, the same motion is imposed on a small
actuation zone at the center of the cylinder~(\cref{fig:visc_stream}b, green).
We achieve this through Brinkman penalization, which models this zone as a
rigid inclusion, allowing us to kinematically \emph{pin} the motion.
The system starts from rest in a stress free state and the simulation is run well beyond
the initial transient phase until steady state rectified streaming patterns emerge.
Further details can be found in the figure caption.

We first characterize the viscous streaming response observed for a rigid cylinder.
Following \citet{stuart1966double}, we characterize streaming response through the streaming Reynolds number
$R_s := V_0^2/ \nu \omega$, based on the oscillatory Stokes boundary layer thickness, also
known as the AC boundary layer thickness $\delta_{AC} := (\nu/ \omega)^{1/2}$, where $\nu$
is the kinematic viscosity of the fluid. \Cref{fig:visc_stream}a shows the time averaged
streamline patterns for this case, depicting the streaming response for $R_s
= 0.63$ ($\delta_{AC} / a = 0.126$), as clockwise (blue) and anti-clockwise (orange)
vortical flow structures around the cylinder.
We note the presence of a well defined boundary layer of thickness $\delta_{DC}$,
also known as the DC boundary layer, commonly used to characterize the topology of streaming flows.
The normalized DC layer thickness $\delta_{DC} / a$ and the AC layer thickness $\delta_{AC} / a$,
can be analytically related as illustrated in \cref{fig:visc_stream}e. As
seen from this figure, our numerical results \cite{parthasarathy2019streaming} compare well with previous boundary
layer scalings based on theory \cite{bertelsen1973nonlinear} and experiments
\cite{lutz2005microscopic}.

Following this rigid body--fluid coupling validation for viscous streaming, we perform a cursory exploration to
observe the effect of cylinder elasticity on the streaming response, by varying the
Ericksen number $\Er := \mu_f V_0 / G a$, where $\mu_f$ is the dynamic viscosity
of the fluid and $G$ is the shear modulus of the cylinder. \Cref{fig:visc_stream}b--d present
the streaming response for increasing values of $Er$, at $R_s = 0.63$ ($\delta_{AC} / a = 0.126$).
When compared to the
rigid body ($Er = 0$) case, the flow structures appear topologically similar, though a decrease
in DC layer thickness $\delta_{DC} / a$ is observed for increasing values of $Er$, or equivalently
with increasing \emph{softness} of the cylinder. Similar trends are observed for different
values of $R_s$ (or $\delta_{AC} / a$) as seen in \cref{fig:visc_stream}e, with the
boundary layer scaling curves becoming less steep (i.e higher deviation from the rigid
body limit) with increasing $Er$. Therefore, perhaps counterintuitively, the strength of
the DC layer increases as the body stiffness decreases, providing a novel avenue for flow
manipulation as well as a potential technique to estimate solid material properties via
flow analysis. A rigorous explanation/rationale for this
behaviour is beyond the scope of the current work and is left as a topic for
future research.

\subsection{Rigid cylinder bouncing on an elastic trampoline}\label{sec:trampoline}
\begin{figure}[!ht]
\centering
\includegraphics[width=0.7\textwidth]{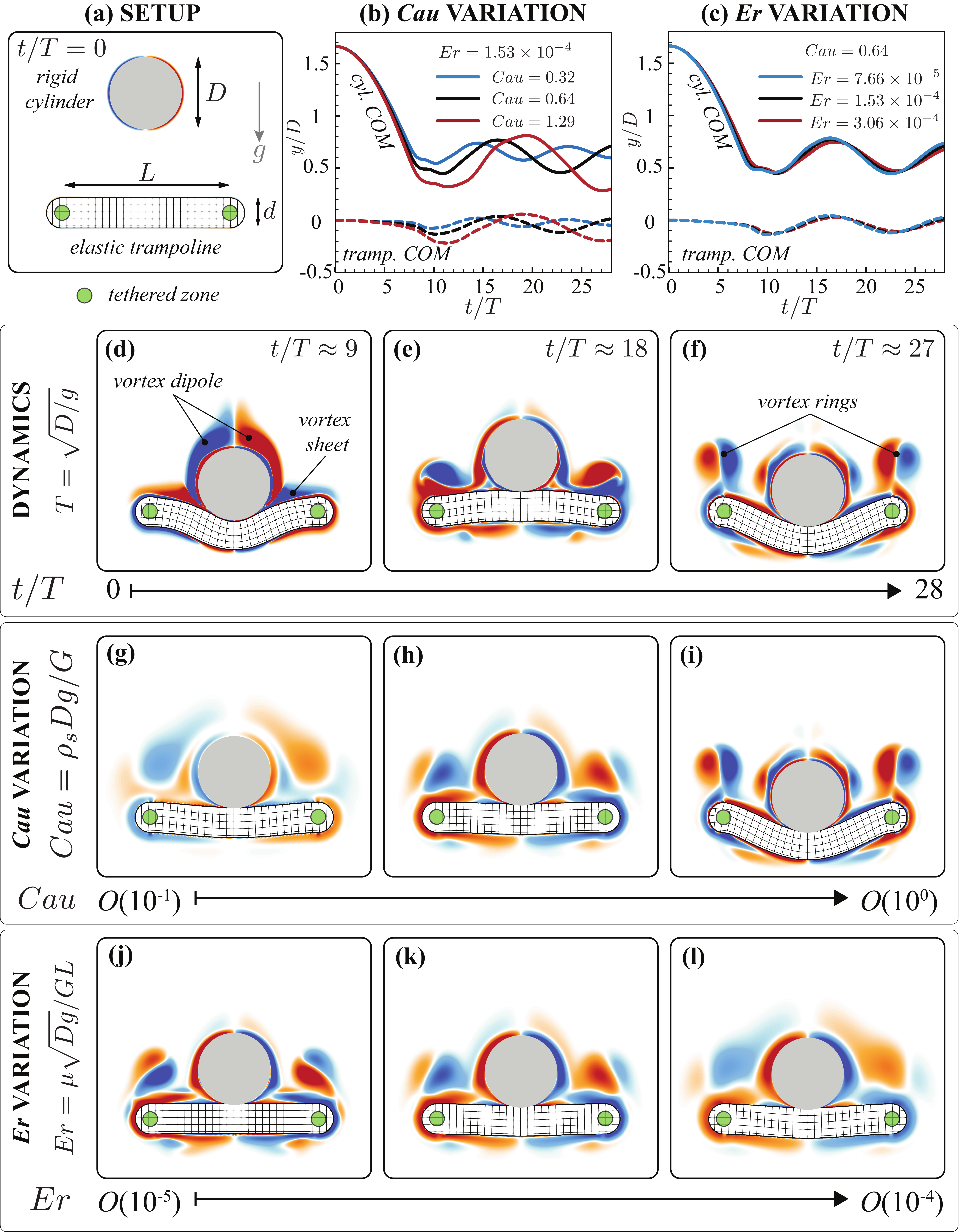}
\caption{
\label{fig:trampoline}
Rigid cylinder--elastic trampoline collision interaction: \capsub{a} setup showing the
initial snapshot. The rigid cylinder with \(\rho_s = 1.1
\rho_{f}\) and diameter \(D = 0.24\) is initialized at \(\left( 0.5, 0.7 \right)\). The details
of the trampoline's exact geometry is deferred to the appendix, here we only
list the important parameters. It has the same density as the cylinder, and is
initialized with centerline at \(y = 0.3\), with length between the anchors \(L = 0.56\) and a thickness \(d = 0.1\), i.e. \(L / d = 5.6\). The tethering
zones are centered at both ends of the trampoline length \(L\) and has a radius
\(r_{\textrm{tether}} = 0.025\) i.e. \(r_{\textrm{tether}} / t = 0.25\).
Both these solids are immersed in a fluid with fixed density \(\rho_{f}\) and a case dependent
dynamic viscosity \(\mu = \mu_{f} \), under a gravity field \(-g \; \hat{\gv{j}}
= -49.05 \; \hat{\gv{j}}\). The trampoline is made from a visco-elastic neo-Hookean material
with case dependent \(G = 2c_1\) and same dynamic viscosity as the fluid \(\mu_s = \mu_{f}\). The
spring stiffness \(k_{\textrm{tether}}\) is set to \(10^{-2}\rho_s
\left(\Delta t\right)^{-2}\) which ensures that the tether's
natural frequency is \(10 \times\) less than the one imposed by the simulation
time-step \(\Delta t\). The tethering \(H\) is mollified by \(\epsilon = 2h\),
where \(h\) is the grid spacing. Furthermore, we enable collision forces
between the solid bodies with \(\epsilon_{\textrm{coll}} = 8h\) and \(k_{\textrm{coll}} = 1 \cdot G\). The key non-dimensional parameters in this case
are \(t / T := t / \sqrt{D / g}\), \(\Ca := \rho_{s} D g / G\) and \(\Er := \mu
\sqrt{D g} / G L\). All simulations are run till \(t / T = 28\). Other
computational parameters are \(h = \left(1024\right)^{-1}, \LCFL = 0.05, \CFL = 0.1\). \capsub{b} Variation of cylinder and
trampoline \(y\) COM with elasticity \(\Ca = 0.32\) (blue), 0.64 (black), 1.29 (red) shows significant
differences compared to \capsub{c} variation with dynamic viscosity \(\Er \cdot 10^{4} = 0.76\) (blue), 1.53 (black),
3.06 (red), thus revealing the importance of \( \Ca \) in determining the system dynamics
within the parameter space investigated here. \capsub{d-f} Depicts the
temporal evolution of vorticity (colored, orange and blue indicate positive and
negative vorticity respectively) and \refmap (black line) contours for a reference case with
\( \Ca = 1.29,\;\Er \cdot 10^{4} = 1.53 \), showing the
deformation of the trampoline and eventual ejection of symmetric vortex rings
due to collision. \capsub{g-i} Shows vorticity and \refmap snapshots at \(t / T
\approx 27\) for varying \(\Ca = 0.32, 0.64, 1.29\) at fixed \(\Er \cdot 10^{4} =
1.53\)---differences can be seen in the trampoline's flexural behavior and vorticity
evolution. \capsub{g-i} Illustrates snapshots at \(t / T \approx 27\) for
varying \(\Er \cdot 10^{4} = 0.76, 1.53, 3.06\) at fixed \(\Ca = 0.64\), where appreciable
differences can be observed only in the vorticity profiles.
}
\end{figure}

We now showcase our solver's ability to capture interactions between
density mismatched rigid and elastic solids of density \(\rho_s \) immersed in a fluid
medium with density \(\rho_{f}\). We
begin, as shown in \cref{fig:trampoline}a, by initializing a dense rigid cylinder
(~\(\rho_{s} = 1.1 \rho_{f}\)) of diameter \(D\) under a gravity field
\(-g \; \hat{\gv{j}}\) in an unbounded domain, at some distance from a
horizontal, dense (~\(\rho_{s} = 1.1 \rho_{f}\)), elastic trampoline of length
\(L\), clamped at the end points.
We clamp the trampoline dynamically using an external body force applied
to the circular tether regions (green zones in \cref{fig:trampoline}a) of the form
\[ \gv{b}(\gv{x}, t) := \gv{f}_{\textrm{tether}}(\gv{x}, t) = k_{\textrm{tether}}
H_{t\epsilon} \left( r_{\textrm{tether}} - r(\gv{x}) \right) ( \gv{x} -
\gv{\xi} \left( \gv{x}, t\right) )\]
which mimics a compact, conservative spring force. Here \(k_{\textrm{tether}}\)
denotes the spring stiffness, H\textsubscript{t\(\epsilon\)} is the tether's mollified
Heaviside function with mollification width \(\epsilon\), \(r_{\textrm{tether}}\) is the
tethering radius and \(r(\gv{x})\) is the radial distance from the tether
point. Additional geometric and parametric details can be found in the figure
caption. This dynamic mode of tethering, in addition to the kinematic mode seen
earlier in \cref{sec:visc_stream}, further illustrates the flexibility of
our solver to account for a variety of boundary conditions. We then let the cylinder fall
and observe the fluid--solid system's response in time (\cref{fig:trampoline}d--f), while
varying (in separate simulations) the trampoline elasticity G, through \(\Ca := \rho_{s} D g / G\)
(\cref{fig:trampoline}g--i), and dynamic viscosity \(\mu = \mu_{f} = \mu_{s} \), through
\(\Er := \mu \sqrt{D g} / G L\) (\cref{fig:trampoline}j--l).

First, we focus on the system dynamics, visualized through vorticity and \refmap
contours in \cref{fig:trampoline}d--f (video provided in the supplementary material).
Here we select a representative set of parameters characterized by $\Er << 1$ and $\Ca
\sim O(1)$. In this scenario, as the cylinder approaches the trampoline, stresses propagate
through the fluid causing the trampoline to deform even before contact takes place.
Concurrently, a vortex sheet at the trampoline surface forms in response to the
cylinder's dipolar vortices and the shear stresses induced by the evacuating
interstitial fluid film (\cref{fig:trampoline}d) \cite{orlandi1990vortex}. Eventually, the fluid
film is entirely squeezed out and the
cylinder collides with and sticks to the trampoline. This in turn causes the
cylinder--trampoline system to start oscillating in the vertical direction(\cref{fig:trampoline}e).
Meanwhile, the trampoline's vortex-sheets and the cylinder's dipolar vorticity merge, laterally ejecting two
symmetric vortex rings (\cref{fig:trampoline}f), which are eventually deflected upwards by
the oscillating trampoline.

We then investigate how this base case scenario varies as a result of changes in $\Ca$,
from a \emph{hard} \(\Ca = 0.32\) (more rigid) to a \emph{soft} \(\Ca = 1.29\) trampoline.
We track the \(y\)-coordinate of the COM of the cylinder and trampoline and report it
in \cref{fig:trampoline}b. The corresponding vorticity and \refmap snapshots
at the final time are reported in \cref{fig:trampoline}g--i. As expected, the
harder trampoline (high G, low \(\Ca\)) does not deform much (blue line, \cref{fig:trampoline}b),
but oscillates at a higher frequency (which we expect from the scaling
\(\omega_{\textrm{osc}} \sim \sqrt{G/\rho_s}/L\)). Instead, as we increase softness,
the trampoline oscillates with smaller frequency but deforms more (red line,
\cref{fig:trampoline}b), which leads to the ejection of the prominent vortex rings seen
in \cref{fig:trampoline}i.

Next, we plot the vorticity and \refmap contours of final time $t/T \approx 27$, as we
vary fluid viscosity from \(\Er \cdot 10^{4} = 0.76\) to \(\Er \cdot 10^{4} = 3.06\)
(\cref{fig:trampoline}j--l). We observe stark differences in the
vorticity contours---as expected, the vortex rings are stronger for a
less viscous fluid and become more diffused as viscosity increases. However
these differences do not affect the system's COM characteristics, which
almost perfectly overlap, as seen from \cref{fig:trampoline}c. We conclude
that within the range of parameters investigated, \( \Ca \) (elasticity)
dominates \( \Er \) (viscosity) in determining the system dynamics.

\subsection{Elastic flag flapping in the wake of a rigid heated cylinder}\label{sec:flag}

\begin{figure}[!ht]
\centering
\includegraphics[width=0.87\textwidth]{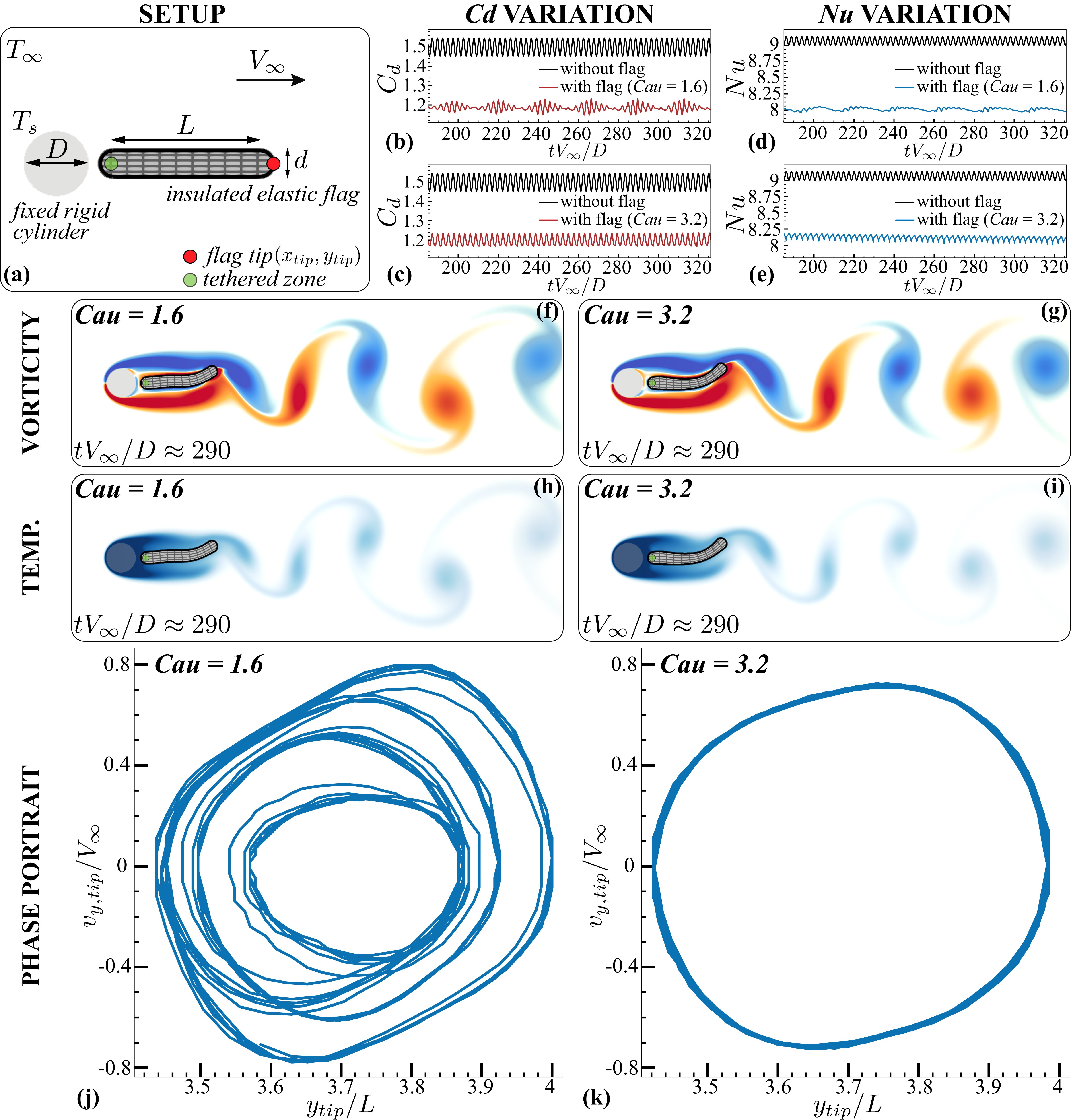}
\caption{
\label{fig:flag}
Elastic flag flapping in the wake of a rigid heated cylinder. \capsub{a} Setup. The
rigid cylinder with diameter \(D
= 0.06\) is initialized at \(\left( 0.1, 0.5 \right)\), and heated to a constant
temperature $T_s = 10$. The details
of the flag's exact geometry is deferred to the appendix, here we only
list the important parameters. It is density matched, and is
initialized at \(y = 0.5\), with length between the anchors \(L = 0.135\),
thickness \(d = 0.024\) (\(L / d = 5.625\)), and is thermally insulated with
initial temperature equaling the ambient temperature $T_{\infty} = 0$. The tethering
zone is centered at the left end of the flag length \(L\) and has a radius
\(r_{\textrm{tether}} = 0.006\) i.e. \(r_{\textrm{tether}} / d = 0.25\).
Both these solids are immersed in a fluid with fixed density \(\rho_{f}\), a case dependent
dynamic viscosity \(\mu = \mu_{f} \) and thermal diffusivity $\alpha$. The fluid has a
background free stream horizontal velocity $V_{\infty} = 4$ and ambient temperature
$T_{\infty}$. The flag is made from a visco-elastic neo-Hookean material with case
dependent \(G = 2c_1\) and same dynamic viscosity as the fluid \(\mu_s = \mu_{f}\).
The spring stiffness \(k_{\textrm{tether}}\) is set to \(10^{-2}\rho_s
\left(\Delta t\right)^{-2}\) which ensures that the tether's natural frequency is 10
times less than the one imposed by the simulation time-step \(\Delta t\). The
tethering \(H\) is mollified by
\(\epsilon = 2h\), where \(h\) is the grid spacing. The key non-dimensional parameters
are \(t V_{\infty} / D \), \(\Ca := \rho_{s} V_{\infty}^2 / G\), \(Re := \rho_f
V_{\infty} D / \mu_f\) and Prandtl number \(Pr := \mu_f / \rho_f \alpha\). For all cases,
we fix $Re = 200$ and $Pr = 1$. All simulations are run till \(t
V_{\infty} / D = 330\). Other computational parameters are \(h = \left(1024\right)^{-1},
\LCFL = \CFL = 0.1\). \capsub{b} Comparison of the temporal drag
coefficient $C_d$ profiles against the baseline no flag case for system with $\Ca = 1.6$ flag and
\capsub{c} $\Ca = 3.2$ flag. \capsub{d} Comparison of the temporal Nusselt number $Nu$ profiles
against the baseline no flag case for $\Ca = 1.6$ flag and \capsub{e} $\Ca = 3.2$ cases.
(f-i) Snapshots of vorticity field (orange and blue indicate positive
and negative vorticity respectively), temperature field (darker shade of blue
corresponds to a higher temperature) and inverse map (black line), for
$t V_{\infty} / D \approx 290$, depicting the flag
flapping motion and vortex shedding for $\Ca = 1.6$ and $\Ca = 3.2$ cases.
Vorticity, carrying \emph{pockets} of high temperature, is shed in a periodic or
quasi-periodic fashion from the cylinder--flag system depending on flag elasticity.
\capsub{j} Phase portrait (i.e tip velocity vs. deflection) of the flag tip
motion plotted over time $t V_{\infty} / D := 180-330$, shown for
$\Ca = 1.6$ and \capsub{k} $\Ca = 3.2$.
}
\end{figure}

Here we demonstrate the multiphysics capabilities of our solver, through the case of
an elastic flag flapping in
the wake of a rigid heated cylinder. Along with dynamical characterization of the flow--structure
interaction, we also characterize the system from a heat transfer perspective.
Additionally, through this case we also highlight one of the important aspects of
remeshed vortex method---relaxed timestep restriction compared to conventional
$\CFL$ bounds.

We begin, as shown in \cref{fig:flag}a, by initializing a fixed rigid cylinder of
diameter $D$ immersed in constant, unbounded, background free stream of velocity
$V_{\infty} \hat{\gv{i}}$. A density matched ($\rho_{e} = \rho$) elastic flag of
length $L$ is initialised at some distance downstream
from the cylinder. We clamp the flag dynamically using a tethering force
(\cref{sec:trampoline}, green zone in \cref{fig:flag}a) at the upstream end, allowing the
flag to flap freely in response to the surrounding flow.
Additional geometric and parametric details can be found in the figure caption.
Following the description of the solid--fluid coupling setup, we then present the setup of
the coupled heat transfer problem and associated governing equations. The cylinder is
maintained at a constant temperature $T_s$ while submerged in a viscous fluid of initial
ambient temperature $T_{\infty}$ and constant thermal diffusivity $\alpha$. The elastic flag is initially at the ambient temperature $T_{\infty}$ but is
thermally insulated and hence does not permit any heat transfer (zero heat flux) across its boundary.
We denote by $\Omega_{r}$ and $\partial \Omega_{r}$ the support and boundary of the
cylinder, while $\Omega_{e}$ and $\partial \Omega_{e}$  stand for the
support and boundary of the flag. The outward normal vector of the flag boundary is
denoted by $\gv{n}$. The temperature field $T(\gv{x}, t)$ is then described by the
scalar advection--diffusion equation with corresponding boundary conditions

\begin{equation}
\label{eqn:temp1}
\begin{alignedat}{3}
    \frac{\partial T}{\partial t} + \left( \gv{v} \cdot \bv{\nabla} \right) T
    &= \alpha \nabla^2 T , ~~~~&&\gv{x}\in\Sigma\setminus\Omega_{r}\setminus\Omega_{e}
    \\
    T &= T_s, ~~~~&&\gv{x}\in \partial \Omega_{r} \\
     \gv{\nabla} T \cdot \gv{n} &= 0, ~~~~&&\gv{x}\in \partial \Omega_{e} \\
     T(\gv{x}, t = 0) &= T_{\infty}, ~~~~&&\gv{x}\in\Sigma\setminus\Omega_{r}
\end{alignedat}
\end{equation}
We solve the governing equations above by extending the penalization technique
for a passive scalar field, similar to the methods described in
\cite{ramiere2007fictitious,kadoch2012volume}, by solving the following modified equations
between steps 42 and 43 of the main algorithm
\begin{equation}
\label{eqn:temp2}
\begin{alignedat}{3}
    \frac{\partial T\pen}{\partial t} + \left[ H(\phi_{r}) \gv{V}_{r} + (1 -
    H(\phi_{r})) ~ \gv{v} \right] \cdot \bv{\nabla} T\pen
    &=
        \lambda H(\phi_{r}) (T_s - T)
        +
        \bv{\nabla} \cdot \left( \left[ \alpha (1 - H(\phi_{e})) +
            \eta\pen H(\phi_{e}) \right]
        \bv{\nabla} T\pen \right)
        ,~~~~\gv{x}\in\Sigma \\
     T\pen(\gv{x}, t = 0) &= T_{\infty}, ~~~~\gv{x}\in\Sigma\setminus\Omega_{r}\\
     T\pen(\gv{x}, t = 0) &= T_{s}, ~~~~\gv{x}\in\Omega_{r}\\
\end{alignedat}
\end{equation}
where $T\pen$, $\gv{V}_r$, $\lambda$, $\eta\pen$, $\phi_r$ and $\phi_e$ correspond to the
penalized temperature field, rigid body velocity of
the cylinder, Brinkman penalization factor, penalized diffusion parameter
(\citet{kadoch2012volume}, set equal to $1e^{-7}$), level set capturing the cylinder boundary
$\partial \Omega_{r}$ and level set capturing the flag boundary $\partial \Omega_{e}$, respectively.
We note that since the cylinder is fixed, $\gv{V}_r = 0$. The Dirichlet condition on the
cylinder (fixed temperature $T_s$) is
imposed via the first term on RHS of~\cref{eqn:temp2}, while the Neumann condition (zero
heat flux) for the flag boundary is achieved by imposing a vanishing diffusivity inside
the flag via the penalized diffusion term \cite{ramiere2007fictitious, kadoch2012volume}
(second term on RHS).
This formulation adds to the flexibility of our solver
by accounting for a variety of boundary conditions from a multiphysics perspective.
Numerically all operators are discretized similar to the
Cauchy momentum equation, described in \cref{sec:num}.

We simulate this cylinder--flag system long enough after shedding vortices to eventually
reach a dynamic, quasi-steady state. This is visualized through vorticity and
temperature contours at a particular time instance, for two flags of different
elasticities $G$, in \cref{fig:flag}f--i. In this state, we characterize
the dynamical and thermal response of the system as functions of flag elasticity $G$
($\Ca = \rho V_{\infty}^2 / G$), by tracking the resulting drag coefficient $C_{d}$ and the Nusselt
number $Nu$

\begin{equation}
\label{eqn:CdNu}
C_{d} := \frac{2 |F_{D, x}|}{D V_{\infty}^2}; ~~~~ Nu := \frac{|Q| D}{(T_s - T_{\infty}) A \alpha}
\end{equation}
where $A$ is the cylinder heat transfer area, $F_{D, x}$ is the horizontal component of
the drag force $\gv{F}_D$ acting on the cylinder, and $Q$ is the heat transfer
rate from the cylinder. We compute these quantities by integrating the penalization term
~\cite{gazzola2011simulations, angot1999penalization, nikhil2018} as shown below

\begin{equation}
\label{eqn:FQ}
\begin{aligned}
    \gv{F}_D = \lambda \int_{\Sigma} H(\phi_r) ( \gv{v}\pen - \gv{V}_r ) ~ d\gv{x} ;
    ~~~~
    Q = \lambda \int_{\Sigma} H(\phi_r) ( T\pen - T_s ) ~ d\gv{x}
\end{aligned}
\end{equation}

We first compare the response seen in these cases to a baseline case in which
the flag is absent.
\Cref{fig:flag}b,c and \cref{fig:flag}d,e present the comparison of temporal $C_{d}$ and $Nu$
profiles for systems with elastic flag corresponding to $\Ca = 1.6$ and $\Ca = 3.2$ (i.e.
stiff vs soft), against the baseline case (which has been validated in \cref{app:no_flag}
against \cite{russell2003cartesian, nakamura2004variation}).
In both cases, we observe a drop of \(\sim 20\%\) in $C_{d}$ and \(\sim 10\% \) in $Nu$
values upon placing an elastic flag in the wake of a cylinder, meaning that the presence
of a flag is favourable in terms of cylinder drag, while detrimental to its heat transfer
properties.
The flag's elasticity \( \Ca\) negligibly alters the values of these diagnostic quantities but
significantly affects their temporal response---while the baseline seems to exhibit
a cyclic sinusoidal behavior, the \emph{soft} ($\Ca = 3.2$) flag induces a cycle asymmetry which gets amplified for a \emph{hard} ($\Ca = 1.6$) flag.
To further investigate the dynamical behavior of the system in these cases,
we temporally track the vertical flapping motion of the flag at its tip location
(red circle in \cref{fig:flag}a) and plot the phase portrait of tip velocity vs
vertical displacement in \cref{fig:flag}j,k. From these plots,
we infer that the \emph{soft} ($\Ca = 3.2$) flag dynamics, reflected as
a limit cycle in the phase portrait, is periodic.
The \emph{hard} ($\Ca = 1.6$) flag's dynamics
is reflected in the phase portrait as a quasi-cycle (an approximate
cycle, that does not repeat exactly), indicating
its quasi-periodic nature. Similar dynamical transitions with variation in elasticity of
flapping flags have been previously reported~\cite{engels2015numerical, turek2011numerical}. Such
variation in dynamics, drag and thermal response, regulated by introducing and varying elasticity, hint towards potential future applications in drag reduction, heat transfer and associated areas.

We conclude this investigation by observing
algorithmic speedups achieved by employing a relaxed $\LCFL$ time step restriction. We report
\(\sim 2 \times\) faster time-to-solutions, compared to a simulation whose time step \(\Delta t \)
is restricted by conventional $\CFL$ criterion. This is consistent with the speedup
expected from the physics of such advection dominated problems. Indeed, for $\Rey \gg 1$
and $\Ca > 1$, the $\Delta t$ restriction due to the free stream $\CFL$
condition 
($\Delta t_{\CFL} = \CFL ~ h / V_{\infty}$)
is more stringent than its counterpart based on the solid shear wave speed ($\Delta t_{sh} = \CFL ~ h
\sqrt{\rho / G}$). Our implementation based on remeshed vortex method can sidestep
this $\Delta t_{\CFL}$ restriction, achieving speed up factors of $\sim \hspace{-0.15cm}\sqrt{\Ca}$.
These considerations further compound the virtues of our approach, on top of its accuracy,
versatility and relative simplicity.

\subsection{Active soft self-propelling swimmers}\label{sec:jellyfish}
\begin{figure}[!h]
\centering
\includegraphics[width=\textwidth]{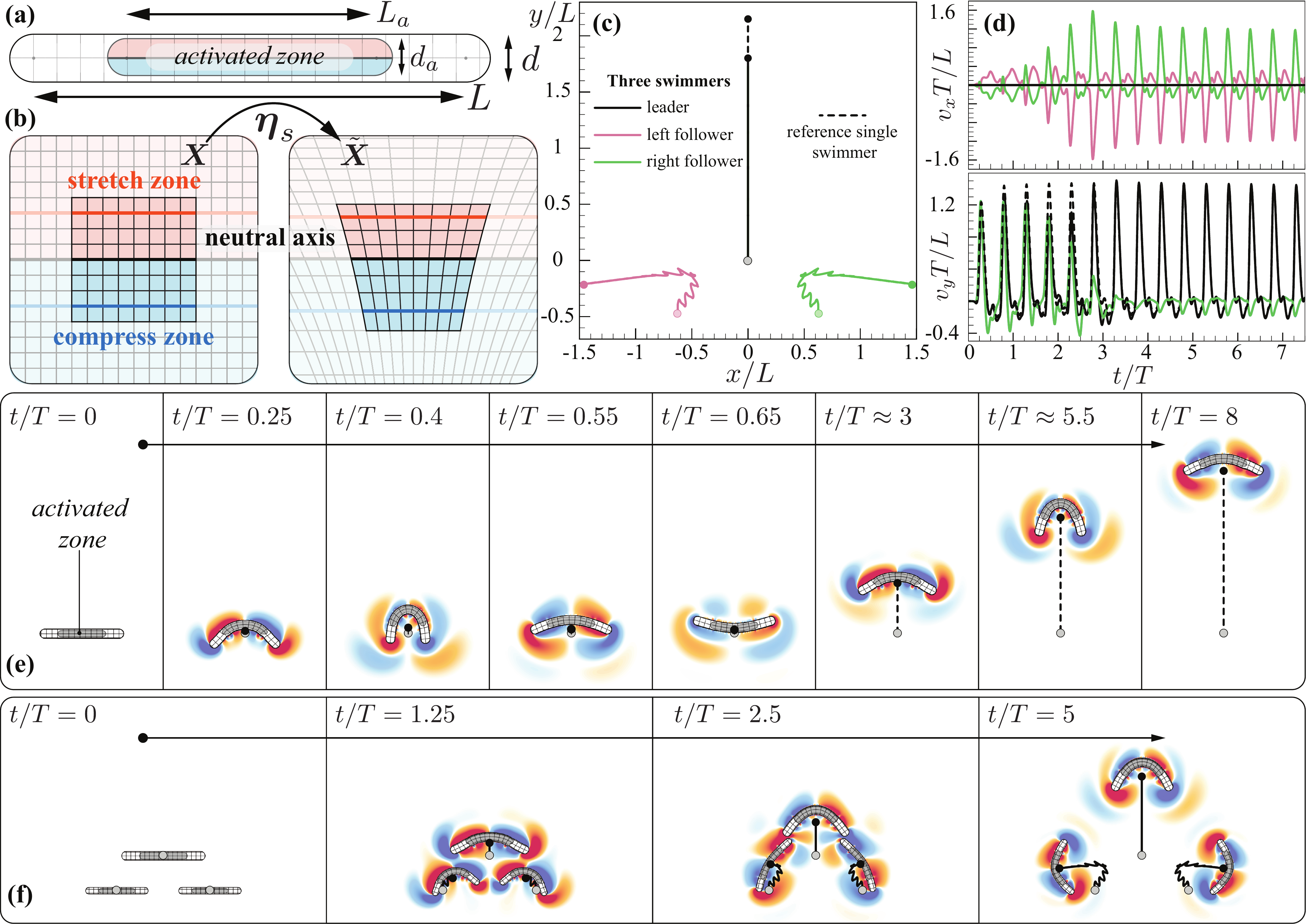}
\caption{
\label{fig:swimmers}
 Single and multiple swimmers \capsub{a} Setup : an active, elastic swimmer of density \(\rho =
1\), elastic modulus \(G = 2.5\) resembling a jellyfish is
immersed in a fluid of density \(\rho = 1\), dynamic viscosity \(\mu = 1e^{-3}\). The details of exact geometry of the swimmer is deferred to the appendix,
here we only list the important parameters. We initialize it with length \(L =
0.2\) and thickness \(d = 0.024\), i.e. \(d / L = 0.12\). The swimmer has
an active region with exactly the same geometry but different parameters \(L_a = 0.112 =
0.56 L\) and \(d_a = 0.0168 =  0.7d\), i.e. \(d_a / L_a = 0.15\). This
active region accommodates a time-periodic incompressible activation map \(\gv{\eta}_s\) of \cref{eq:eta_swimmer} with \(\lambda_s\) set to \(\log(2.2) / (d_a / 2)\) which indicates a maximum elemental
stretch/compression of \(2.2 \times\) (see below) and \(\omega = 2 \pi / 3.125\). This results in the following non-dimensional parameter set \(\Ca :=
\rho\omega^2L^2 / G = 0.064\) and \(\Er := \mu \omega / G = 8.04 \cdot 10^{-4}\).
\capsub{b} Schematic presenting the effects of this map \(\gv{\eta}_s\) on a
regular Cartesian grid. We first demarcate the centerline neutral axis (solid black) upon
which the map has no effect. Above the neutral axis resides the \emph{stretching} (red) zone
where elements are stretched horizontally while being squished vertically to
maintain incompressibility. The amount of stretch \(\lambda_s\) is reflected
in the degree of stretch of a material line element marked above in solid red.
Simultaneously, below the neutral axis lies the \emph{compression} (blue) zone, where
elements are compressed horizontally. Applying this map time-periodically, while
preserving polarities of the stretch and compress zones, results in alternate bending and
relaxing of the centerline leading to a propulsive motion of the swimmer. The resulting non-dimensional
trajectories of single (dashed black line) and multiple (solid lines) swimmers
is shown in \capsub{c}, where the initial and final positions are marked with
hollow and solid circles respectively. \capsub{d} Non-dimensional horizontal velocity
and \capsub{e} non-dimensional vertical velocity of the swimmers as they evolve with time. \capsub{f}
Time-series of snapshots of a single, vertically locomoting swimmer and its grey active
region along with vorticity (colored, orange and blue indicate positive and negative
vorticity respectively) and inverse map (black line) contours. For reference,
the swimmer is initialized at \((0.5, 0.35)\) and is allowed to move till \(t / T =
8\). Additionally, the center of mass location with time is highlighted by a dashed black line.
\capsub{g} Showcases a similar time-series, but this time for three similarly
activated elastic swimmers---one adult leader and two juvenile followers. The leader has the
exact geometry and proportions of the single swimmer of (f) and is initialized
at the same location \((0.5, 0.35)\). The juveniles have the same geometry as the leader but are
\(75 \%\) its size. They are initialized symmetrically with the left follower
at \((0.375, 0.25625)\) and right follower at \((0.625, 0.25625)\). In this
case, we also enabled collision forces between the bodies with \(\epsilon_{\textrm{coll}} = 8h\) and \(k_{\textrm{coll}} = G\). The center of mass histories of the leader, left follower and right follower are marked with a solid black, pink and green lines respectively. The snapshots show the followers
closing in towards the symmetry axis, \emph{kissing} the
leader and making a hard, right turn to then proceed almost horizontally for the
rest of the time. Meanwhile, the leader continues on its vertical upward path.
Other pertinent computational parameters are \(h = \left(1024\right)^{-1}, \LCFL = 0.05, \CFL =
0.1\).
}
\end{figure}

Finally, in our last demonstration we showcase the ability of our method to seamlessly
incorporate endogenous muscular actuation, a feature of importance in bio-locomotion and
biophysical settings, through the example of self-propelling swimmers. We consider single and multiple
density matched (\(\rho_{f} = \rho_{s} = \rho\)) elastic swimmers of dimension \(L\), resembling the two-dimensional cross-section of a
jellyfish, submerged in fluid with dynamic viscosity \(\mu\) as shown in \cref{fig:swimmers}(a).
Within each of these swimmers,
we have an activated region that mimics the action of localized, internal muscles. We
utilize the formulation of \cref{sec:invmap} and actuate this region by using the following
time-periodic activation map \(\gv{\eta}_s\)
\begin{equation}
\gv{\eta}_s
\left(
\gv{X} :=
\begin{pmatrix}
X \\ Y
\end{pmatrix}, t
 \right) :=
\begin{pmatrix}
Xe^{\lambda_{s}Y H_{\varepsilon}(\phi_a) \sin^8 \left(\omega t \right)}\\
(1 - e^{-\lambda_{s}Y H_{\varepsilon}(\phi_a) \sin^8 \left(\omega t \right)}) / \lambda_{s}\\
\end{pmatrix}
\label{eq:eta_swimmer}
\end{equation}
where \(H_{\varepsilon}(\phi_a)\) is the indicator function of the active region, meant to
localize the effects of \(\gv{\eta}_s\) and \(\omega\) is the angular
frequency. The symbol \(\lambda_{s}\) here indicates a stretch factor---indeed \(\gv{\eta}_s\) stretches and compresses elements away from the swimmer
centerline, while maintaining incompressibility (i.e. \(\det{\left( \bv{\nabla}\gv{\eta}_s
\right) } \equiv 1\), see \cref{fig:swimmers}(b) and corresponding caption).
The surrounding unactivated
solid region is passive and responds to the effects of the activation above. The elastic
modulus of both the activated and unactivated regions is denoted by \(G\). We
note that similar setups were investigated before qualitatively~\cite{zhao2008fixed,
rycroft2018reference}, but not quantitatively. Here, we complement previous studies with a
rigorous quantitative characterization, for reproducibility.

We begin by observing the locomotion of a single swimmer of length \(L\), for a
representative case with \(\Ca := \rho\omega^2L^2 / G = 0.064\) and \(\Er := \mu
\omega / G = 8.04 \cdot 10^{-4}\). The swimmer flaps
its appendages and moves upward, causing the generation and shedding of trailing
vortices as shown in \cref{fig:swimmers}(e). The region of activation is highlighted in
gray. We track the swimmer COM coordinates and
velocities and report them in \cref{fig:swimmers}(c) and \cref{fig:swimmers}(d),
respectively.
These plots indicate that the swimmer follows a perfectly vertical trajectory due to symmetry.
We observe the \(\Rey := {V}_{\textrm{max}} L / \nu\) to be \(\approx 20\),
based on the maximum velocity \({V}_{\textrm{max}}\) during the course of the
swimmer's trajectory. We note that even though the actuation is periodically symmetric,
the resulting forward speed is periodically asymmetric with noticeable accelerations during the
power stroke. This break in temporal symmetry, which helps propel the swimmer faster, arises due to
elastic relaxation time scales pervasive throughout the swimmer body. The motion then
emerges due to a complex interplay between actuation, elasticity and morphology, whose parametric
details can be found in the figure caption.

Next, we place three swimmers in a triangular formulation, with one adult
\emph{leader} jellyfish and two juvenile \emph{followers}. The leader has the same
proportions as the single swimmer in the simulations above. Both
followers are scaled down versions of the leader (parametric details
can be found in the figure caption). We activate each of these swimmers similar to
the previous case. In this case, the flow-mediated collective behavior leads to complex
dynamics as seen from~\cref{fig:swimmers}(g). We first focus on the followers. Their
trajectory is significantly
affected by the vorticity shed by the leader. They are first drawn closer
together towards the symmetry axis, shortly after which they closely approach
and \emph{kiss} the leader's
appendages. Due to this near-approach event, they make a near-perpendicular \(90^o\)
turn and continue propelling in the horizontal direction. Meanwhile the
leader persists on its expected straight, vertical trajectory, seemingly
unaffected by the followers. On a closer comparison with the trajectories of an equivalent
single swimmer (i.e. without the followers) simulation in~\cref{fig:swimmers}(c), we see that it is
slowed down. The rationale for this behavior is uncovered from the vertical velocity \(v_y\) plots of~\cref{fig:swimmers}(d).
For \( t/T < 3 \) the followers, which are in close proximity to the leader, slow it down (solid black line vs dashed black for a single swimmer) . Once the leader frees itself from the followers' influence, it swims with the same speed as the single swimmer, seen for \( t/T > 4 \).

Finally, we draw attention to the preservation of symmetry in the swimmer trajectories (in
\cref{fig:swimmers}(d)) and the flow fields of \cref{fig:swimmers}(g) even after long times and a critical
near-approach event. This, along with a battery of tests and illustrations, attests to the accuracy and robustness of our FSI approach.

\section{Conclusion}\label{sec:conc}

In conclusion, we have presented a unified framework based on remeshed vortex method
for the simulation of mixed rigid/elastic bodies immersed in a viscous fluid. Our approach
seamlessly incorporates a rigid body-fluid interaction formulation based on Brinkman
penalization and projection, within a broader elastic body-fluid methodology based on
inverse map technique and one continuum formulation. Our formulation produces a neat
relatively simple algorithm, whose accuracy and
robustness is demonstrated through rigorous benchmarking and convergence analysis,
against a battery of theoretical/numerical tests. Through various multifaceted
illustrations (which themselves may serve as detailed benchmarks for future studies), we
further demonstrate our solver's versatility, applicability and robustness across multiphysics
scenarios, boundary conditions, constitutive and actuation models, along
with algorithmic speedup for advection dominated problems. In
particular, the broad range of physics captured involving muscular actuation, multi-body
contact, self propulsion and heat transfer illustrates the utility of our method in a
range of applications, from bio-locomotion to heat transfer and microfluidics.
The use of particle methods and simple
convenient grid based operators renders the solver scalable and makes it portable to
parallel architecture such as GPUs and multicores \cite{rossinelli2015mrag}.
Accordingly, the development of a HPC implementation able to take advantage of modern
heterogeneous computing infrastructures to simulate 3D and/or thousands of immersed
elastic/rigid bodies in realistic physical time, remains in the scope of future work.



\section{Acknowledgements}

The authors acknowledge support by the National Science Foundation under NSF
CAREER Grant No. CBET-1846752 (MG) and by the Blue Waters project
(OCI- 0725070, ACI- 1238993), a joint effort of the University of Illinois
at Urbana-Champaign and its National Center for Supercomputing Applications.
This work used the Extreme Science and Engineering Discovery Environment (XSEDE)
\cite{towns2014xsede} Stampede2, supported by National Science Foundation grant number
ACI-1548562, at the Texas Advanced Computing Center (TACC) through allocation
TG-MCB190004.


\appendix

\section{Oscillatory response in parallel layers of fluid and solid:
Density mismatch validation}\label{app:rho_mismatch}

\begin{figure}[!ht]
    \centering
    \includegraphics[width=0.7\textwidth]{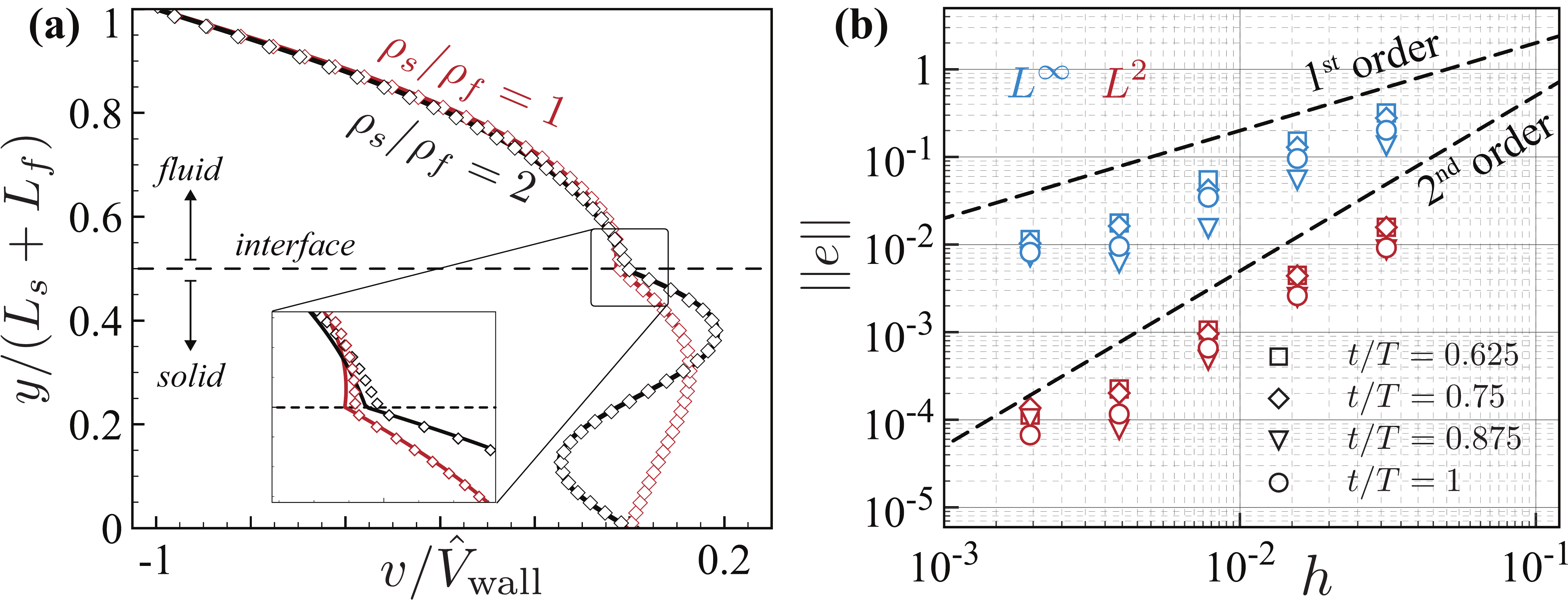}
    \caption{
    Oscillatory response in parallel density mismatched solid--fluid layers for a
     neo-Hookean visco-elastic solid. We retain the physical setup of
    \cref{sec:bmk2} and run simulations to obtain the velocity fields at the
    center of the domain, shown in \capsub{a} (black), which once again agrees
    with the analytical results. We also plot the velocity traces
    of the corresponding density matched case from the main text in red to contrast it
    with this case---indeed stark differences are seen within the solid phase.
    The inset shows the
    concentration of errors near the diffuse interface. For reference, numerical
    results are plotted with scatter points whereas
    analytical results are plotted with a solid line.
    Tracking these velocity
    results with changing resolution results in the convergence plot shown in
    \capsub{b} where \Linf (blue) and \Ltwo (red) norms of the error are plotted
    against grid spacing $h$ at different \( t / T\). Trends indicate a first
    to second order convergence as expected.
    The dynamic parameters corresponding to the fluid phase in this setup are
    \( \rho_{f} = 1, \mu_{f} = 0.02\). The dynamic parameters of the elastic
    solid  are \( \rho_{e} = 2 \), \(\mu_{e} = 0.1\mu_{f} \) and shear modulus
    \( c_1 = 0.01\) i.e. \( G = 2c_1 = 0.02\). The simulations are run
    until \(t / T = 10 \), and physical quantities are sampled within the last cycle.
    The key non-dimensional dynamic parameters for this benchmark are
    \( \Rey = {\rho_{f} \dot{\gamma} L_{f}} / {\mu_{f}} = 2 ,
    \Er = {\mu_{f} \hat{V}_{\textrm{wall}} } / {2 G L_{s}} = 1\). The computational
    parameters are set to \( \LCFL = 0.05, \CFL = 0.1\).    }
    \label{fig:bmk2_rho_mismatch}
\end{figure}

Here, we showcase our ability to accurately capture effects of density mismatch
by once again comparing numerical results to analytical ones in the case of oscillating,
parallel sandwiched elastic solid--fluid layers. We retain the physical setup and
explanation of~\cref{sec:bmk2}, and focus on the results for the density mismatch case
\( \rho_{e} = 2, \rho_{f} = 1\). These
results are presented in~\cref{fig:bmk2_rho_mismatch}(a) (black) and contrasted to the
density matched results from the main text (red). We clearly see stark differences
in the velocity profiles within the solid phase. Once again, maximum differences
between analytical and numerical results are seen in the diffuse interface region.
Plotting convergence by retaining the corresponding definition of error used in
the main text for different temporal instants in ~\cref{fig:bmk2_rho_mismatch}(b)
reveals consistent first to second order convergence (1.86 for \Ltwo and 1.19 for
\(\Linf\)), as expected.

\section{Convergence of incompressibility errors inside the solid}\label{app:mass_convg}
\begin{figure}[!ht]
\centering
\includegraphics[width=0.68\textwidth]{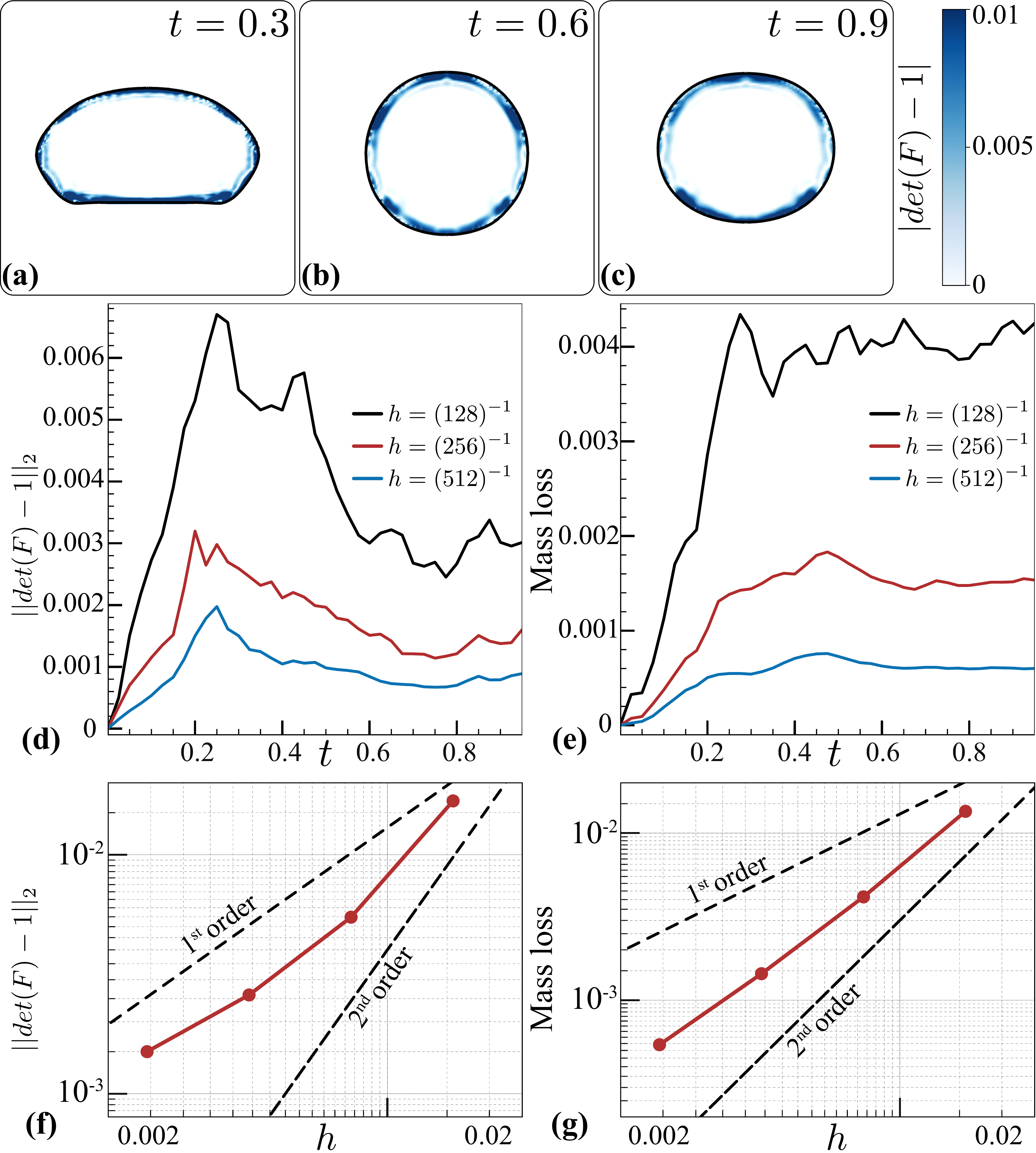}
\caption{
    \label{fig:detF_loss}
    Convergence of incompressibility errors inside the solid. \capsub{a-c} Temporal
    variation of the $|{det}(\bv{F}) - 1|$ field contours (colored,
    darker shade of blue represents higher values) for one of the solids from the
    benchmark illustrated in \cref{sec:bmk4}. Localisation of the error can be seen in the
    solid--fluid blur/blending zone, near the interface (black contour). Computational
    details can be found in \cref{fig:bmk4} caption. Temporal variation of \capsub{d}
    \Ltwo norm of $|{det}(\bv{F}) - 1|$ field,  $||{det}(\bv{F})
    - 1||_2$ and \capsub{e} mass loss plotted against time $t$, for different spatial
    resolutions. Spatial convergence: \capsub{f} $||{det}(\bv{F}) - 1||_2$
    and \capsub{g} mass loss plotted against grid spacing $h$, at time $t = 0.3$.
}
\end{figure}

For incompressible elastic solids with density $\rho_e$, incompressibility dictates that
the mass of a differential element of the solid should be constant, and therefore locally the determinant of the
deformation gradient ${det}(\bv{F})$ should be identically equal to 1. As shown in
\citet{jain2019conservative}, an incompressible velocity field ensures that the above
conditions are satisfied, although only in the continuous limit but not necessarily in the discrete
limit. We note that a consistent, accurate numerical implementation then produces incompressibility
errors that are bounded and convergent. We then demonstrate this robustness and
accuracy by presenting the quantification and convergence of incompressibility errors inside the
solid, encountered in our method, in the benchmark case of collision between two
hyperelastic cylinders immersed in a fluid, illustrated in \cref{sec:bmk4}.

\Cref{fig:detF_loss}a-c showcase the qualitative temporal variation of incompressibility errors for
one of the solids, captured through contours of $|{det}(\bv{F}) - 1|$ field. We observe
that this field is bounded and
localised in the solid--fluid blur/blending zone around the interface (black contour),
at all times. Additionally, incompressibility is
ensured within the pure solid zone at all times. In
order to quantify and demonstrate convergence for these incompressibility errors we compute two diagnostic quantities following
\cite{jain2019conservative, rycroft2018reference}. These are  the \Ltwo norm of the
$|{det}(\bv{F}) - 1|$ field, and the total mass loss of the solid computed as

\begin{equation}
    \label{eqn:mass_loss}
    \textrm{Mass loss} = 1 - \frac{M_e(t)}{M_e(t = 0)}
    = 1 - \frac{\rho_e \int_{\Sigma} H(\phi_e(t)) ~ d\gv{x}}
    {\rho_e \int_{\Sigma} H(\phi_e(t = 0)) ~ d\gv{x}}
\end{equation}
where $M_e$ represents the total mass of the solid and $\phi_e$ is the level set capturing the interface of the body.
\Cref{fig:detF_loss}d,e present the temporal variation of
$||{det}(\bv{F}) - 1||_2$ and mass loss, respectively, at different spatial
resolutions. We observe that errors increase with deformation, i.e reach their highest
values at maximal deformation ($t \approx 0.3$), and then decrease again or saturate
with time to a nearly constant value, hence showing no accumulation of errors over time in
the  present approach. Additionally, both these diagnostic quantities are seen to converge with spatial resolution. We present this
spatial convergence in
\cref{fig:detF_loss}f, g, retaining the computational parameters of
\cref{sec:bmk4}. As seen from this figure, the convergence order for both diagnostics was found to be between first and
second order (least squares fit of 1.2 for $||{det}(\bv{F}) - 1||_2$ and 1.5
for mass loss), which is consistent with the spatial discretization of our solver.
Thus, our solver is consistent and accurate in ensuring
incompressibility in both the solid and fluid phases.

\section{Geometrical details of trampoline, flag and swimmers}\label{app:geometry}
The trampoline (\cref{sec:trampoline}), flag (\cref{sec:flag}) and swimmers (\cref{sec:jellyfish})
shown in the main text are constructed using the same geometry which we now
discuss. The geometry is essentially a rounded rectangle---made of a central
rectangle with two semi-circles at its ends---with the diameter of the end circles
matching the width of the central rectangle. The geometry is characterized by its
left center point \( \gv{x}_c := (x_c, y_c)^T\), the length (\(L\)) and thickness (\(d\)) of the
rectangle. Its level set function is then described by

\[
\phi
\left(
\gv{x} := (x, y)^T
\right) :=
\begin{cases}
\sqrt{(x-x_c)^2 + (y - y_c)^2} - d / 2 &\mbox{for } x < x_c \\
\lvert y - y_c \rvert - d / 2 &\mbox{for } x_c \leq x \leq x_c + L \\
\sqrt{(x-x_c-L)^2 + (y - y_c)^2} - d / 2 &\mbox{for } x > x_c + L
\end{cases}
\]

\section{Dynamic and thermal validation for flow past a cylinder}\label{app:no_flag}

Here, we briefly present the validation for the no flag variant (i.e flow past a rigid
cylinder) for the illustration case of elastic flag flapping in the wake of a rigid heated
cylinder, described in \cref{sec:flag}. In order to validate the dynamical and thermal response,
we present a comparison of commonly used diagnostic quantities, which include the mean drag
coefficient $\overline{C}_d$, mean Nusselt number $\overline{Nu}$ and the Strouhal number
$S\hspace{-0.05cm}t$, against previously
published results \cite{russell2003cartesian, nakamura2004variation}. We compute
$\overline{C}_d$ and $\overline{Nu}$ based on
\cref{eqn:CdNu}, while $S\hspace{-0.05cm}t$ is computed as follows

\begin{equation}
    S\hspace{-0.05cm}t = \frac{f D}{V_{\infty}}
\end{equation}
where $f$, $D$ and $V_{\infty}$ correspond to the vortex shedding frequency, cylinder
diameter and free stream velocity, respectively. \Cref{no_flag_table} shows the comparison
of the values of the above quantities obtained using our method against those found in
previous works for Reynolds number $\Rey = 200$ and Prandtl number $Pr = 1$. We
note that our results show close agreement with the previously published values.
For a more detailed validation of this case, the reader is referred to our previous work~\cite{gazzola2011simulations}.

\begin{table}[htbp]
\centering
\begin{tabular}{cccc}
\toprule
& $\overline{C}_d$ & $\overline{Nu}$  & $S\hspace{-0.05cm}t$\\
\midrule
Previous results & 1.45 & 9.05 & 0.20\\
Present methods & 1.49 & 9.06 & 0.19\\
\bottomrule
\end{tabular}
\caption{
\label{no_flag_table}
Dynamic and thermal validation for flow past a cylinder. Comparison of drag
coefficient $\overline{C}_d$, Nusselt number $\overline{Nu}$ and the Strouhal
number $S\hspace{-0.05cm}t$ computed with the present
method against previously published results \cite{russell2003cartesian,
nakamura2004variation} at Reynolds number $Re = 200$ and Prandtl number $Pr = 1$.
For computational details, refer to \cref{fig:flag} caption.}
\end{table}

\bibliographystyle{elsarticle-num-names}
\bibliography{cfs_lit.bib}





\end{document}